%% file: main.tex
\documentclass{aa}

\usepackage{natbib}
\usepackage{amsmath}
\bibpunct{(}{)}{;}{a}{}{,} 
\usepackage{graphics}   
\usepackage{graphicx}   
\usepackage{epstopdf}
\usepackage{longtable}   
\usepackage{url}      
\usepackage{bm}        
\usepackage[varg]{txfonts}
\usepackage{pdflscape}
\usepackage{supertabular}
\usepackage{hyperref}   
\usepackage[svgnames]{xcolor}
\usepackage{longtable}
\usepackage{lscape}
\usepackage{booktabs}
\usepackage{graphicx}
\usepackage{multicol}
\usepackage{placeins}
\usepackage{afterpage}

\usepackage[T1]{fontenc}
\usepackage{lipsum}
\usepackage[normalem]{ulem}

\usepackage{float}
\usepackage{svg}
\usepackage[shortlabels]{enumitem}

\newcommand{\nga}{($n,\gamma$) }
\newcommand{\rms}{\mathrm{rms}}

\begin{document}

\title{The intermediate neutron capture process}
\subtitle{IV. Impact of nuclear model and parameter uncertainties}
\author{S\'ebastien Martinet\inst{1}, Arthur Choplin\inst{1}, Stephane Goriely\inst{1}, Lionel Siess\inst{1}}

 \institute{Institut d’Astronomie et d’Astrophysique, Universit\'e Libre de Bruxelles (ULB), CP 226, B-1050 Brussels, Belgium\\
 email: sebastien.martinet@ulb.be}
 \authorrunning{Martinet et al.}

\date{Accepted: October 11, 2023}

\abstract
{The observed surface abundance distributions of Carbon-enhanced metal-poor (CEMP) r/s-stars suggest that these stars could have been polluted by an intermediate neutron-capture process (the so-called i-process) occurring at intermediate neutron densities between the r- and s-processes. Triggered by the ingestion of protons inside a convective He-burning zone, the i-process could be hosted in several sites, a promising one being the early AGB phase of low-mass low-metallicity stars. The i-process remains however affected by many uncertainties including those of nuclear origin since it involves hundreds of nuclei for which reaction rates have not yet been determined experimentally.}
{We investigate both the systematic and statistical uncertainties associated with theoretical nuclear reaction rates of relevance during the i-process and explore their impact on the i-process elemental production, and subsequently on the surface enrichment, for a low-mass low-metallicity star during the early AGB phase.}
{We use the TALYS reaction code (Koning et al. 2023) to estimate both the model and parameter uncertainties affecting the photon strength function and the nuclear level densities, hence the radiative neutron capture rates. The impact of correlated systematic uncertainties is estimated by considering different nuclear models, as detailed in Goriely et al. (2021). In contrast, the uncorrelated uncertainties associated with local variation of model parameters are estimated using a variant of the backward-forward Monte Carlo method to constrain the parameter changes to experimentally known cross sections before propagating them consistently to the neutron capture rates. 
The STAREVOL code (Siess et al. 2006) is used to determine the impact of nuclear uncertainties on the i-process nucleosynthesis in a 1 $M_{\odot}$ [Fe/H] = - 2.5 model star during the proton ingestion event in the early AGB phase. A large nuclear network of 1160 species coherently coupled to the transport processes is solved to follow the i-process nucleosynthesis.} {We find that the non-correlated parameter uncertainties lead the surface abundances uncertainties of element with $Z\geq 40$ to range between 0.5 and 1.0 dex, with odd-$Z$ elements displaying higher uncertainties.
The correlated model uncertainties are of the same order of magnitude, and both model and parameter uncertainties have an important impact on potential observable tracers such as Eu and La.
We find around 125 important \nga reactions impacting the surface abundances, including 28 reactions that have medium to high impact on the surface abundance of elements that are taken as observable tracers of i-process nucleosynthesis in CEMP stars. }  
{Both the correlated model and uncorrelated parameter uncertainties need to be estimated coherently before being propagated to astrophysical observables through multi-zone stellar evolution models. Many reactions are found to affect the i-process predictions and will require improve nuclear models guided by experimental constraints. Priority should be given to the reactions influencing the observable tracers.}

\keywords{Nuclear reactions, Nucleosynthesis, Abundances - Stars: AGB and post-AGB}


\maketitle

\section{Introduction}

Most of the elements heavier than Fe are synthesized by the slow (s) and rapid (r) neutron capture processes \citep[e.g.][for a review]{arnould20}. These processes are characterized by neutron densities of $N_n=10^5 - 10^{10}$ and $N_n>10^{24}$~cm$^{-3}$ respectively. The s-process develops in Asymptotic Giant Branch stars (AGB) \citep[e.g.][]{gallino98,herwig05,cristallo11,karakas14, Goriely18c} and in the helium-burning core of massive stars \citep[e.g.][]{langer89,prantzos90,choplin18}. The r-process requires explosive conditions and could arise during the merging of two neutron stars \citep[e.g.][]{arnould07, goriely11a, goriely11b, wanajo14, just15}, in collapsars or magnetorotational supernovae \citep{winteler12,nishimura15,siegel19}.
It is believed that other secondary neutron-capture processes also exist {such as the} n-process, with neutron densities of typically $10^{18}$~cm$^{-3}$, that can develop in the helium-burning shell of massive stars during core-collapse supernovae \citep{blake76,thielemann79,meyer04,choplin20}. The isotopic composition of meteoritic grains may bear the signatures of this process \citep{meyer00,pignatari18}.

The so-called intermediate neutron capture process \cite[or i-process, first named by][]{cowan77} is another secondary neutron-capture process that may develop in a variety of astrophysical sites \citep[see][for a detailed list]{choplin21} and in particular includes low-metallicity low-mass AGB stars \citep[e.g.][]{iwamoto04,cristallo09b,suda10,choplin21,choplin22,goriely21,gilpons22}. 
For the i-process to develop, protons must be mixed in a convective helium-burning zone. This event is often called proton ingestion event (PIE). In AGB stars, PIEs can develop during the early Thermally-Pulsing (TP) phase. It occurs when the energy of the convective thermal pulse is high enough to overcome the entropy barrier at the bottom of the H-burning shell. The top of the convective pulse encroaches {on} the H-shell and protons are engulfed in the pulse. They are transported downwards by convection (in a timescale of typically 1~hr) and burn on the way via $^{12}$C($p,\gamma$)$^{13}$N. After the decay of $^{13}$N to $^{13}$C (in about 10~min), the $^{13}$C($\alpha,n$)$^{16}$O reaction is activated, mostly at the bottom of the convective pulse, where the temperature reaches $\sim 250$~MK. The neutron density goes up to $\sim 10^{15}$~cm$^{-3}$ and an i-process nucleosynthesis takes place. Quickly after the peak in neutron density, the convective pulse splits \citep[cf. Sect.~3.5 in][for a discussion about the split]{choplin22}. The upper part eventually merges with the large convective envelope, in which the i-process products are diluted and finally expelled by the stellar wind.

More and more stars  are observed showing chemical compositions compatible with i-process nucleosynthesis. It includes the so-called Carbon-Enhanced Metal-Poor (CEMP) r/s-stars \citep{lugaro12,roederer16,karinkuzhi21,mashonkina23,hansen23}. Some {less metal-poor} stars may also bear the signature of an i-process \citep{mishenina15,karinkuzhi23}. It has also been suggested that some pre-solar grains could be made of i-process material \citep{fujiya13,jadhav13,liu14}.

Despite the {occurrence of PIE} in various astrophysical sites, such as the early AGB-phase of low-mass low-metallicity stars \citep[e.g.][]{choplin22} or the rapidly accreting white dwarfs \citep[e.g.][]{denissenkov19}, nucleosynthesis predictions are still significantly affected by nuclear uncertainties. These uncertainties were discussed in a few previous works. More specifically, 
\cite{denissenkov18} investigated the uncertainties affecting elements with $35 \leq Z \leq 40$ using both a one-zone model with constant temperature and density and a 1D multi-zone stellar model where nucleosynthesis is calculated in post-processing. They randomly varied 52 relevant \nga reaction rates of unstable species. Because one-zone models led to significantly different results that multi-zone stellar models, they suggested that one-zone models are not reliable for identifying critical reaction rates in convective-reactive regimes such as AGB star experiencing i-process.
Similarly, \cite{mckay20} considered the uncertainties for elements with $32 \leq Z \leq 48$ using one-zone modeling and varying 113 relevant \nga rates of unstable species to predict the impact of (n,$\gamma$) reaction rate uncertainties on the abundances of i-process elements in observed metal-poor stars. 
More recently, \cite{denissenkov21} examined the uncertainties on elements with $56 \leq Z \leq 74$ with both one-zone and multi-zone models. They randomly varied 164 relevant \nga rates of unstable species. By contrast to \cite{denissenkov18}, they suggested that one-zone simulations are reliable to identify key reaction rates provided that the neutron density is similar to the maximum one found in the multi-zone stellar models.
\cite{goriely21} evaluated the impact of nuclear model uncertainties on the surface abundances of 1D multi-zone AGB models for all elements. Nuclear uncertainties related to $^{13}$C($\alpha,n$)$^{16}$O reaction rate, as well as experimentally unknown $\beta$-decay and radiative neutron capture rates were studied. The direct capture contribution, the photon strength functions (PSFs) and nuclear level densities (NLDs) entering the calculation of \nga rates were shown to be the main source of nuclear uncertainties.

In this paper, we study both nuclear model (or equivalently ``systematic'') and nuclear parameter (often referred to as ``statistical'') uncertainties affecting the prediction of theoretical radiative neutron capture rates and their impact on the i-process for elements with $14 \leq  Z \leq 92$. We varied the 868 theoretical \nga rates included in our i-process reaction network and considered a low-mass low-metallicity AGB star model to analyse the impact of nuclear uncertainties on the prediction of elemental and isotopic surface enrichments.
{Section 2 presents the method to obtain model and parameter uncertainties, with a special emphasis on the application of the Backward-Forward Monte Carlo (BFMC) approach to constrain parameter uncertainties on experimental data}. In Sect. 3, we study the impact of both {the parameter and model} uncertainties on the i-process nucleosynthesis in our low-mass low-metallicity AGB star. In Sect. 4, we present the \nga reactions impacting most the surface abundances of such stars, and we {illustrate how the use of a newly constrained rate can significantly reduce the uncertainties}. Finally in Sect. 5, we discuss the results of this work and the potential perspectives of this sensitivity study.


\section{Method}
\label{sec:meth}

As shown in \citet{goriely21}, the main source of nuclear uncertainties impacting the abundance predictions by the i-process is found in the theoretical determination of \nga rates for neutron-rich nuclei. In contrast to our previous analysis \citep{goriely21} and other works dedicated to this subject \citep{denissenkov18,mckay20,denissenkov21}, both the model (systematic) and parameter (statistical) uncertainties affecting theoretical \nga rates are studied here. They are both obtained using the TALYS reaction code \citep{Koning23}. Since most of the atomic masses are known for nuclei produced by the i-process \citep{Wang21}, the key nuclear ingredients affecting the calculation of the radiative neutron capture are the NLDs and PSFs\footnote{Note that, due to its still complex modelling, the direct contribution to the reaction mechanism is neglected in the present analysis \citep[see][]{goriely21}.}. 
{The radiative neutron capture cross section $\sigma_{n,\gamma}$ also formally depends on the neutron-nucleus optical model potential. However, for nuclei produced by the i-process, the cross section remains insensitive to the optical potential due to the prevalence the strong interaction with respect to the electromagnetic one. Indeed, within the Hauser-Feshbach formalism, $\sigma_{n,\gamma} \propto T_n T_\gamma / (T_n+T_\gamma) \sim T_\gamma$, if the neutron transmission coefficient $T_n$ is significantly larger than the electromagnetic one $T_\gamma$, {\it i.e.} when $T_n>>T_\gamma$.  For exotic neutron-rich nuclei, the optical potential may affect the cross section, provided the isovector imaginary potential becomes significant, as discussed in \citet{Goriely07}. {This is not the case for the i-process nucleosynthesis} and the neutron-capture cross section can, in a very good approximation, be assumed to remain essentially {insensitive to} the optical potential. 
}
The nuclear uncertainties are consequently directly related to our ability to estimate NLDs and PSFs 
for the compound systems produced during the i-process. Those are detailed below.

\subsection{Nuclear model uncertainties}
\label{sec:mod}

{The model uncertainties are treated in a similar way as done in \citet{goriely21} by estimating the 868 rates with different combinations of NLD and PSF models. More specifically, we adopt here 9 different combinations based on the following 
{\it i)} NLD models: 
\begin{enumerate}[(1)]
\item Hartree-Fock-Bogolyubov plus combinatorial (HFB+comb) \citep{Goriely08b}
\item Constant-Temperature plus Fermi Gas (Cst-T) \citep{Koning08}
\item Back-Shifted Fermi Gas model (BSFG) \citep{Koning08}
\item T-dependent HFB plus combinatorial (THFB+comb) \citep{Hilaire12}
\end{enumerate}
and {\it ii)} PSF models: 
\begin{enumerate}[(a)]
\item Gogny-HFB plus quasi-particle random phase approximation (D1M+QRPA) \citep{Goriely18a}
\item Simple Modified Lorentzian (SMLO) \citep{Goriely18b}
\item Generalized Lorentzian (GLO) \citep{Kopecky90}
\item Skyrme-HFB plus QRPA (BSk27+QRPA) \citep{Xu21}
\item Relativistic mean-field + continuum RPA (RMF+cRPA) \citep{Daoutidis12a}
\end{enumerate}}
The 9 combinations of NLDs (from 1 to 4) and PSFs (from a to e) are defined as the following: set A (1a), set B (2b), set C (2a), set D (1c), set E (1e), set F (1d), set G (3a), set H (4a) and set I (1b).
These combinations all lead to a relatively accurate prediction of the experimental Maxwellian-averaged cross sections (MACS) \citep{Dillmann06} for all the 239 nuclei with $20 \le Z \le 83$. To quantify this accuracy, we adopt the root-mean-square (rms) criterion defined by the $f_\rms$ deviation, {\it i.e.} 
\begin{equation}
    {f_\rms}=\exp\left[ {\frac{1}{N_e}\sum^{N_e}_{i=1}\ln^2 \left(\frac{\langle \sigma_{\mathrm{th},i}\rangle}{\langle \sigma_{\exp,i}\rangle}\right)}~\right]^{1/2}
\label{eq_frms}
\end{equation}
where $N_e$ is the number of known reaction rates, and $\langle \sigma_{\rm th}\rangle$ and $\langle \sigma_{\rm exp}\rangle$, the theoretical and experimentally known MACS, respectively. We find that for the 9 adopted combinations, $f_\rms$ deviations range between 1.4 and 1.8. Note, however, that in these TALYS calculations, all NLDs are constrained on measured resonance s-wave spacings and low-lying scheme of excited levels, when available \citep{Capote09}, but no information on experimental average radiative width is included.

\begin{figure}
    \centering
    \includegraphics[trim={2.0cm 2.0cm 1.0cm 1.5cm},width=0.45\textwidth]{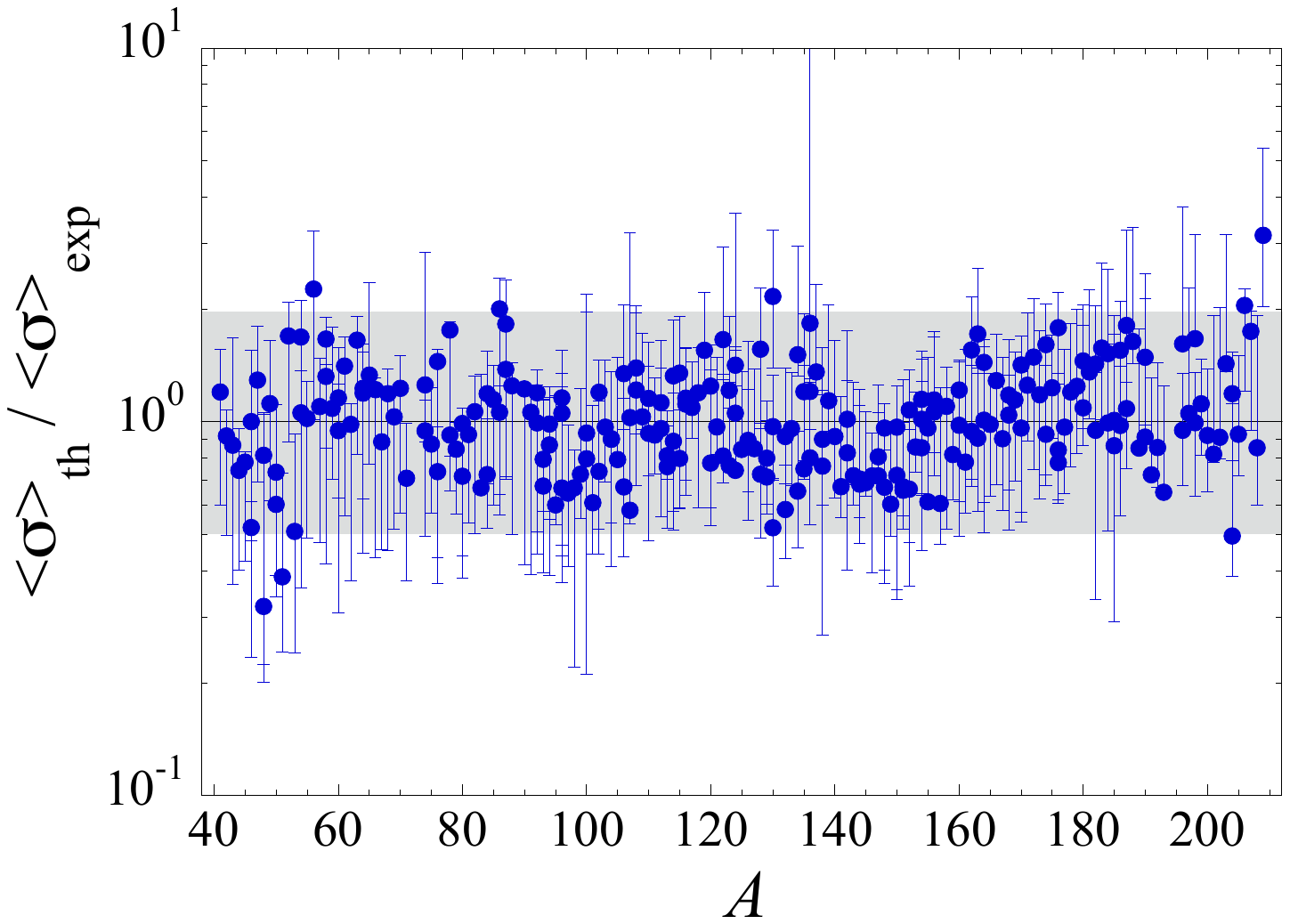}
    \caption{Theoretical over experimental \nga MACS for the 239 experimentally known rates at 30~keV. The error bars display the maximum and the minimum rates obtained by the 9 different combinations of NLD and PSF models (including sets A and B, see text). The full circles correspond to set B and the grey shaded area shows variations within a factor of 2. }
    \label{fig:fig_sigv_exp_model}
\end{figure}

{For each of the $N_e=239$ nuclei for which data is available, TALYS deviations to the experiment is displayed in Fig.~\ref{fig:fig_sigv_exp_model} where the error bars give the upper or lower MACS obtained by one of the 9 combinations. It should be {stressed} that the corresponding model uncertainties obtained with the 9 combinations of NLDs and PSFs are strongly correlated by the underlying nuclear model.}

{For the neutron-rich nuclei relevant to the r-process, for which no experimental information of any kind is available, model uncertainties have been shown to dominate over the parameter uncertainties \citep[see in particular][who discuss the extrapolation uncertainties of mass models]{Goriely14}. However, when dealing with the unstable nuclei close to the valley of $\beta$-stability produced by the i-process, parameter uncertainties may be significant. Those are estimated below.
}

\subsection{Nuclear parameter uncertainties}
\label{sec:par}

{In comparison with model uncertainties, much less effort has been devoted to estimate the parameter uncertainties for a given set of nuclear reaction rates. These are obtained by local variations of the parameters used in a given nuclear model. {For this study we consider} two different combinations of NLD and PSF models for the TALYS calculation of the neutron capture rates. The first set A adopts the HFB+comb NLDs \citep{Goriely08b} and the D1M+QRPA PSFs for both the dipole electric E1 and magnetic M1 components \citep{Goriely18a}. While set A is based on rather microscopic ingredients, set B considers more phenomenological models, namely Cst-T NLDs \citep{Koning08}  and SMLO PSFs \citep{Goriely18b}.
Both sets A and B are sensitive to the parameters adopted in the NLD and PSF models, each of them being  dominantly adjusted by two parameters. For the Cst-T NLDs, we allow for local variations of the temperature $T$ and the pairing parameter $E_0$ leading to a possible energy shift. In the case of the tabulated HFB+comb NLDs, equivalently, two parameters $\alpha$ and $\delta$ can play a similar role, as detailed in \citet{Goriely08b}. For the SMLO and D1M+QRPA PSF models, uncertainties affecting the width and centroid energy of the E1 giant dipole resonance are included by adjusting two related parameters, denoted here as $\delta_\Gamma$ and $\delta_E$, respectively. All together, for both sets A and B, local variations of four parameters (two for NLDs and two for PSFs) may consequently affect the reaction rate predictions. However, the range of local variation for each parameter remains to be defined and will be critical to estimate the magnitude of their impact on the predicted reaction rates. For this reason, it is fundamental that such local parameter variations be constrained as much as possible by experimental data where available, before being applied to neutron-rich nuclei for which no data is available. In our case, it is possible to estimate the impact of such parameter uncertainties on calculated rates by propagating them by Monte Carlo (MC) sampling constrained by available experimental rates \citep{Dillmann06}. The method used here corresponds to the BFMC approach, as detailed below.
}

\subsubsection{The Backward-Forward Monte Carlo approach}
\label{sect:BFMC}
The BFMC method \citep[see][]{Chadwick2007,Bauge2011,Goriely14} relies on the sampling of the model parameters and the
use of a generalized $\chi^2$ estimator to quantify the likelihood of each {simulation} respective to a given set of experimental constraints, here the experimentally known \nga rates \citep{Dillmann06}. The backward MC is used {to select} the suitable parameter samples that agree with experimental constraints. 
{In the BFMC method, the $\chi^2$ estimator is used to {quantify} a likelihood function that will
weight a given sample of  \{$p_1$,...,$p_n$\} parameters when the $N_e$ associated
observables \{$\sigma_1$,...,$\sigma_{N_e}$\} are close to the experimental data.} In this work, we assume that experimentally constrained \nga rates are independent. This assumption allows us to use a $\chi^2$ criterion instead of a generalized weighting function, so the weighting function becomes simply 1 when $\chi^2 \leq \chi_{\rm crit}^2$, and 0 otherwise, with $\chi_{\rm crit}^2$ a chosen critical value of the $\chi^2$.
Then, for the forward MC step, the selected set of the MC parameter sample is applied to the calculation of the unmeasured quantities, here the unknown \nga rates. 

\subsubsection{Assessing TALYS parameter uncertainties}

To estimate the uncertainty affecting our 4 model parameters, we use as $\chi^2$ estimator, the $f_\rms$ deviation with respect to the 239
experimental \nga MACS at 30 keV from the KADoNiS database \citep{Dillmann06}. 
Each of the 4 parameters $p$ affecting
the NLDs and PSFs have been varied separately and assigned a relative uncertainty $\Delta p/p$ {such} that an individual
change of each parameter leads to a maximum 10\% increase
of the $f_\rms$ deviation with respect to the 239 experimental
nuclear rates in our sample. 
{The diagonal values of a covariance matrix is then assigned the values of $\Delta p$ corresponding to the 10\% increase.}
This matrix is used to produce a multivariate normal distribution centered on the nominal parameters values. We {then generate} a distribution of random combinations (here $N_{\rm comb}=200$ combinations) of these 4 parameters. 

Note that the 10\% increase in the $f_\rms$ deviation is chosen to
optimize the sampling needed to get a significant statistics in
the BFMC method, but is not expected to affect the
results. Indeed, a value larger than 10\% would essentially request more
combinations to be calculated to achieve a similar representative sample
of runs constrained by the $\chi^2$ estimator on experimental data.

\subsubsection{Backward Monte Carlo step}
\label{Sect:backward_MC}
\begin{figure}
    \centering
    \includegraphics[width=0.48\textwidth]{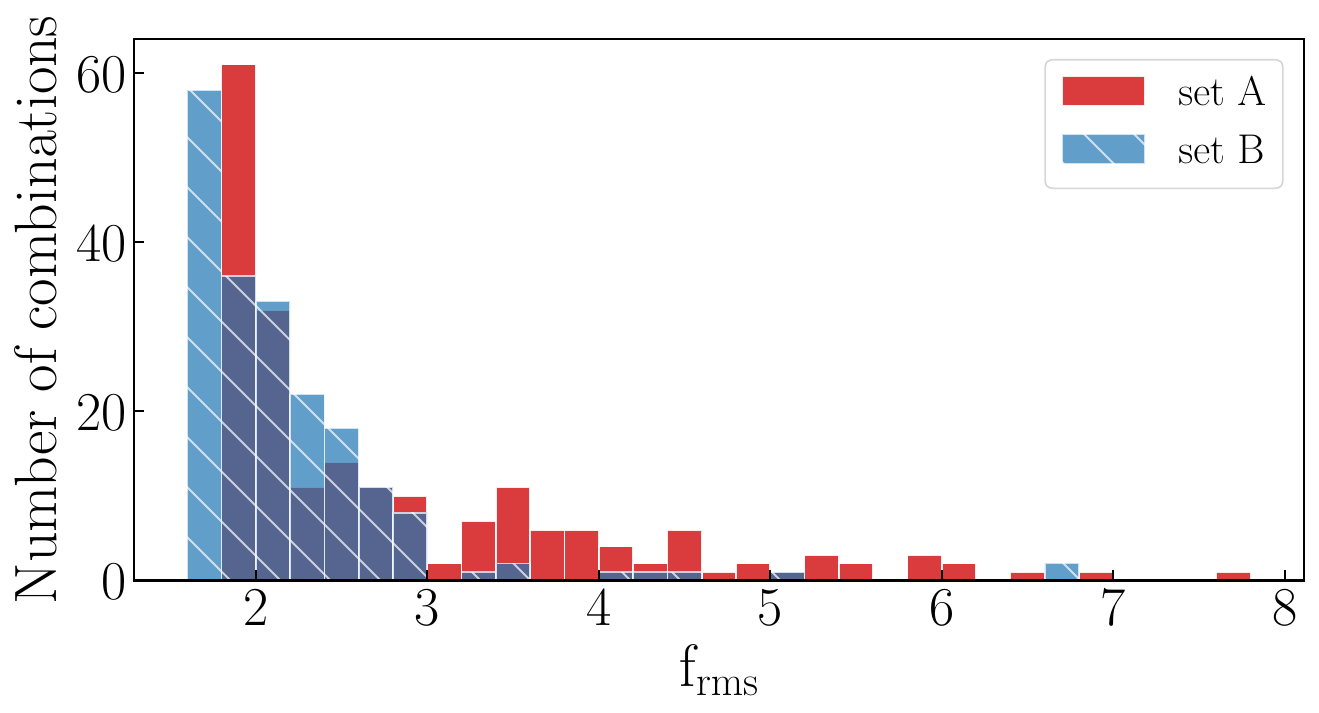}
    \caption{Histogram of the number of parameter combinations for sets A and B as a function of their $f_\rms$ with respect to the experimental MACS.}
    \label{fig:Histogram_rms}
\end{figure}
\begin{figure}
    \centering
    \includegraphics[trim={2.0cm 1.0cm 2.0cm 2.0cm},width=0.40\textwidth]{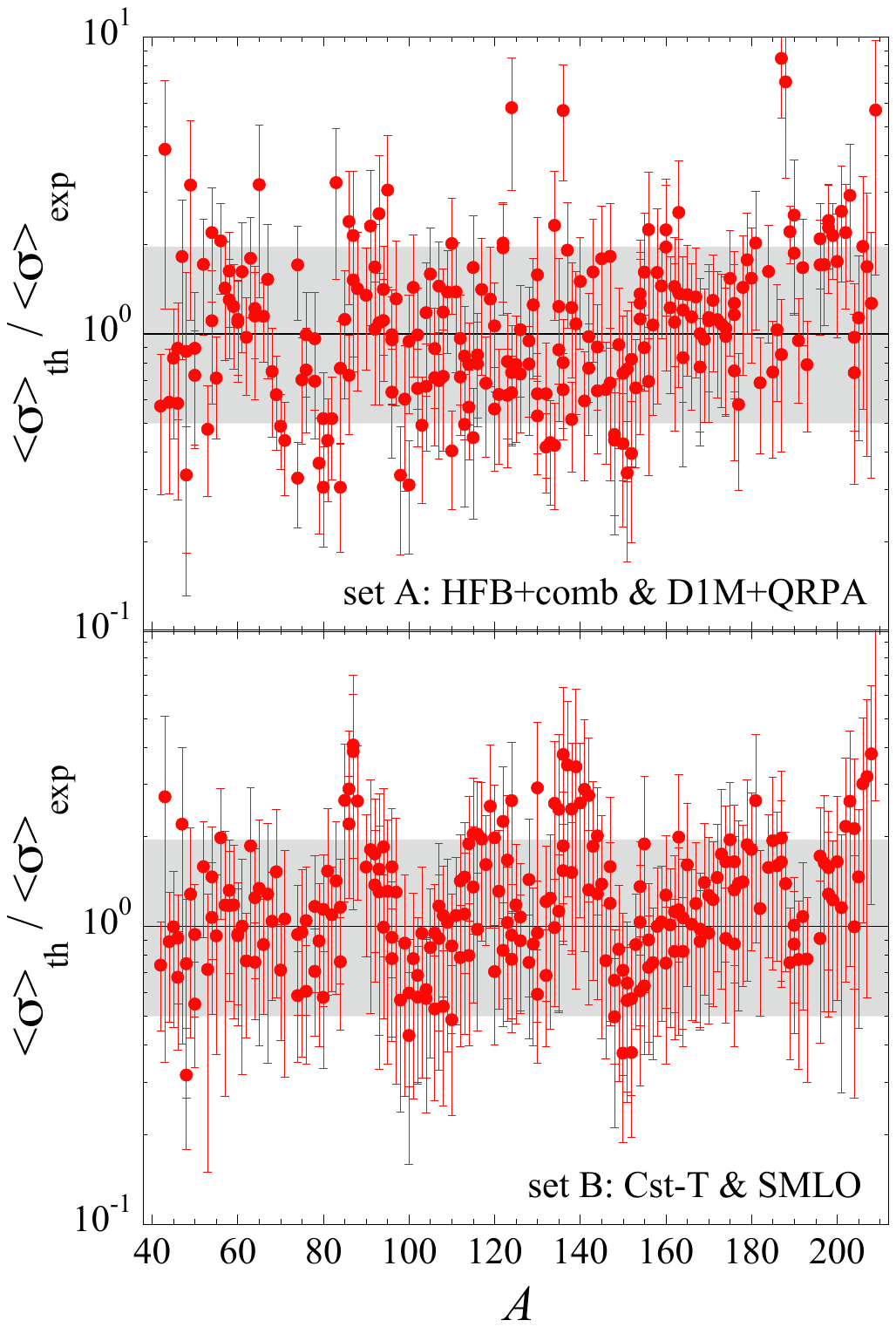}
    \caption{Theoretical over experimental \nga MACS for the 239 ($20\le Z \le 83$) known rates at 30~keV obtained with set A (top) and set B (bottom). The error bars display the uncorrelated parameter uncertainties obtained through the BFMC method using $f_\rms \leq$ 2.0. The grey shaded area guide the eye for a deviation within a factor of 2.}
    \label{fig:fig_sigv_exp_parameter}
\end{figure}

We computed the 239 theoretical rates for which experimental data is available for each of the 200 different combinations of parameter values obtained from the multivariate normal distribution described above.
The $\chi^2$ estimator of the backward MC step is again given by the $f_\rms$ deviation (Eq.~\ref{eq_frms}) obtained with respect to the 239 experimental \nga rates. 
The resulting distribution as a function of the $f_\rms$ deviation is displayed in Fig. \ref{fig:Histogram_rms}. 
{We find no combination resulting in $f_\rms$ deviations smaller than 1.80 for set A and 1.59 for set B. However, as seen in Fig. \ref{fig:Histogram_rms}, many parameter combinations lead to 
$f_\rms$ deviations larger than typically 2. Such combinations lead to an unrealistic description of experimental data and should consequently not be considered for the unknown nuclei during the forward MC step. {Since the model deviations obtained in Sect.~\ref{sec:mod} all give rise to $f_\rms\le 2$ with respect to experimental data, the parameter ranges in the backward MC procedure are also restricted to variations compatible with an $f_\rms\le 2$. This allows us to select only the combinations of nuclear parameters that are consistent with the experimental data by using the $\chi^2 \leq \chi_{\rm crit}^2$ selective criterion with $\chi_{\rm crit}^2 \Leftrightarrow f_\rms\le 2$.}
The resulting selection of model parameters allows us to estimate the uncorrelated parameter uncertainties for the 239 experimentally known MACS, as illustrated in Fig. \ref{fig:fig_sigv_exp_parameter} for both sets A and B.  Some systematic effects can be observed in the parameter uncertainties, especially for set B, for which the MACS is overestimated for closed-shell nuclei around $A\sim 90$, 140 and 208, mainly due to the rather approximate treatment of shell effects in the Cst-T NLD formula. These uncertainties are also seen to be of the same order of magnitude as those stemming from model uncertainties shown in Fig.~\ref{fig:fig_sigv_exp_model}.
Now that our parameter sample is constrained on experimental data, it can be extended to the non-experimental \nga rates by the forward step of the MC procedure, as explained below.
}



\subsubsection{Forward Monte Carlo step}

\begin{figure}
    \centering
    \includegraphics[trim={2.0cm 1.0cm 2.0cm 2.0cm},width=0.40\textwidth]{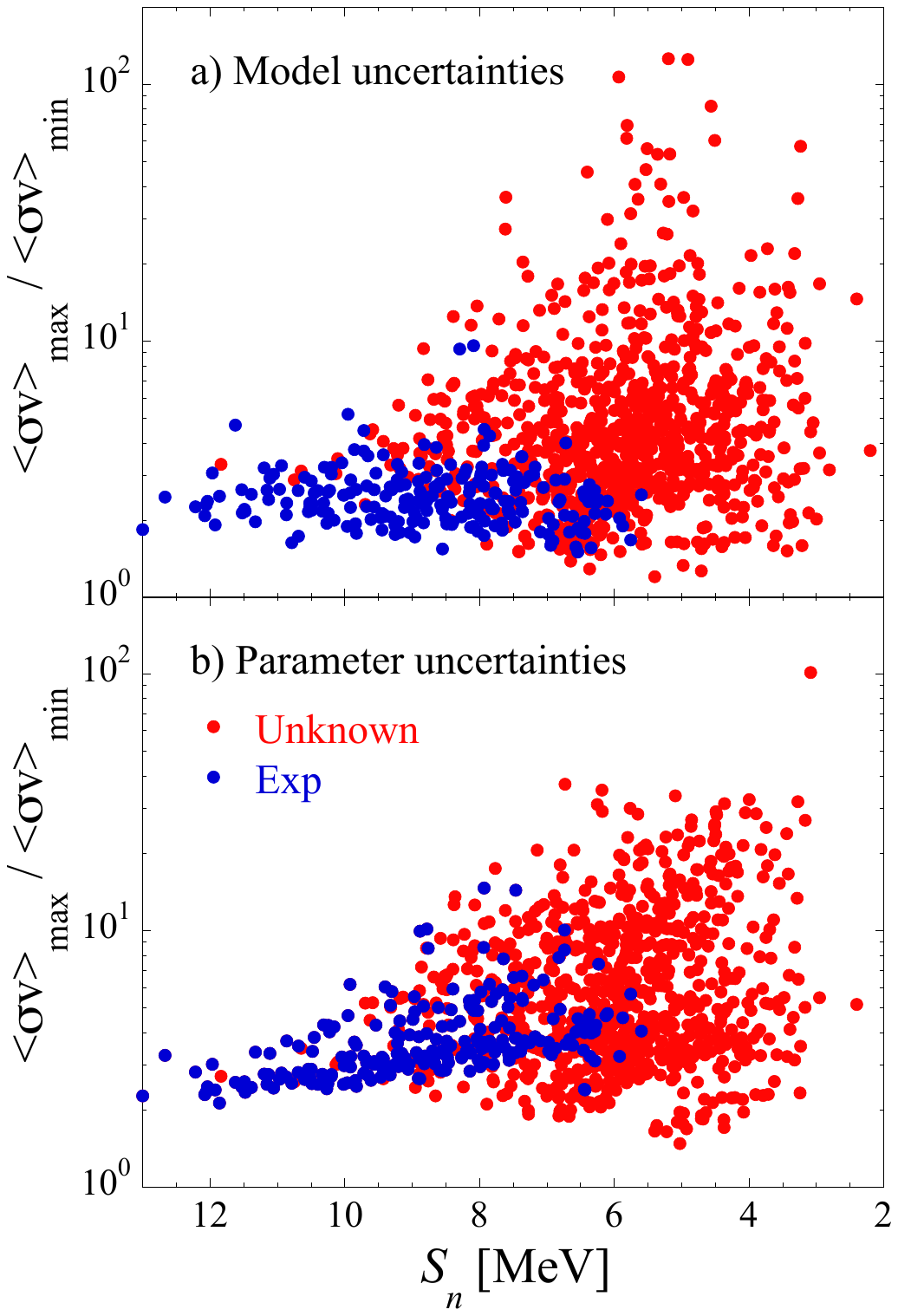}
    \caption{Uncertainties as a function of the neutron separation energy $S_n$. {\it a)} Model uncertainties between sets A and B represented by the ratio of the maximum to minimum \nga rates (max$\left[ \langle\sigma \rangle_A,\langle\sigma \rangle_B\right]/$min$\left[\langle\sigma \rangle_A,\langle\sigma \rangle_B\right]$). The 239 experimentally known reactions are shown with blue dots and the unknown ones by red dots. 
    {\it b)} Same as the upper panel but for the parameter uncertainties for set B. All rates are estimated at $T=2.5 \times 10^8$~K. }
    \label{fig:parameters_uncertainties_Sn}
\end{figure}

\label{sect:forward_MC}
{At the end of the backward MC step, we obtain} a subset of $N^\xi_{\rm comb}$ combinations of parameters constrained by experimental MACS ($f_\rms\le 2$). With this subset we can now assess the non-correlated parameter uncertainties affecting the theoretical \nga rates.
We computed the 868 experimentally unknown \nga rates using the subset of parameter combinations ($N^\xi_{\rm comb}=61$ combinations for set A and $N^\xi_{\rm comb}=97$ for set B). 
The resulting parameter uncertainties on the rates are plotted in Fig. \ref{fig:parameters_uncertainties_Sn} (lower panel) by displaying for each reaction the ratio between the maximum and the minimum MACS obtained using set B (we obtain similar results for set A). The 239 experimentally known rates are shown in blue and the theoretical rates in red. Additionally, for comparison, we show in the upper panel of Fig. \ref{fig:parameters_uncertainties_Sn} the maximum-to-minimum rate ratios obtained between the sets A and B with their nominal parametrisation, {\it i.e.} max$\left[ \langle\sigma \rangle_A,\langle\sigma \rangle_B\right]/$min$\left[\langle\sigma \rangle_A,\langle\sigma \rangle_B\right]$, corresponding to the model uncertainties between sets A and B. Deviations due to both the correlated model and the non-correlated parameter uncertainties are seen to be of the same order of magnitude and are found to increase with decreasing neutron separation energy $S_n$ (\textit{i.e} for increasingly neutron-rich nuclei). Interestingly, for both types of uncertainties, we still find nuclei with low deviations (typically lower than a factor of 2-3) at low $S_n$ thanks to their relatively well known {scheme of excited states} which reduces the impact of NLDs in these cases.

\subsection{Parameter correlations}

\begin{figure}
    \centering
    \includegraphics[trim={2.0cm 1.0cm 2.0cm 1.0cm},width=0.45\textwidth]{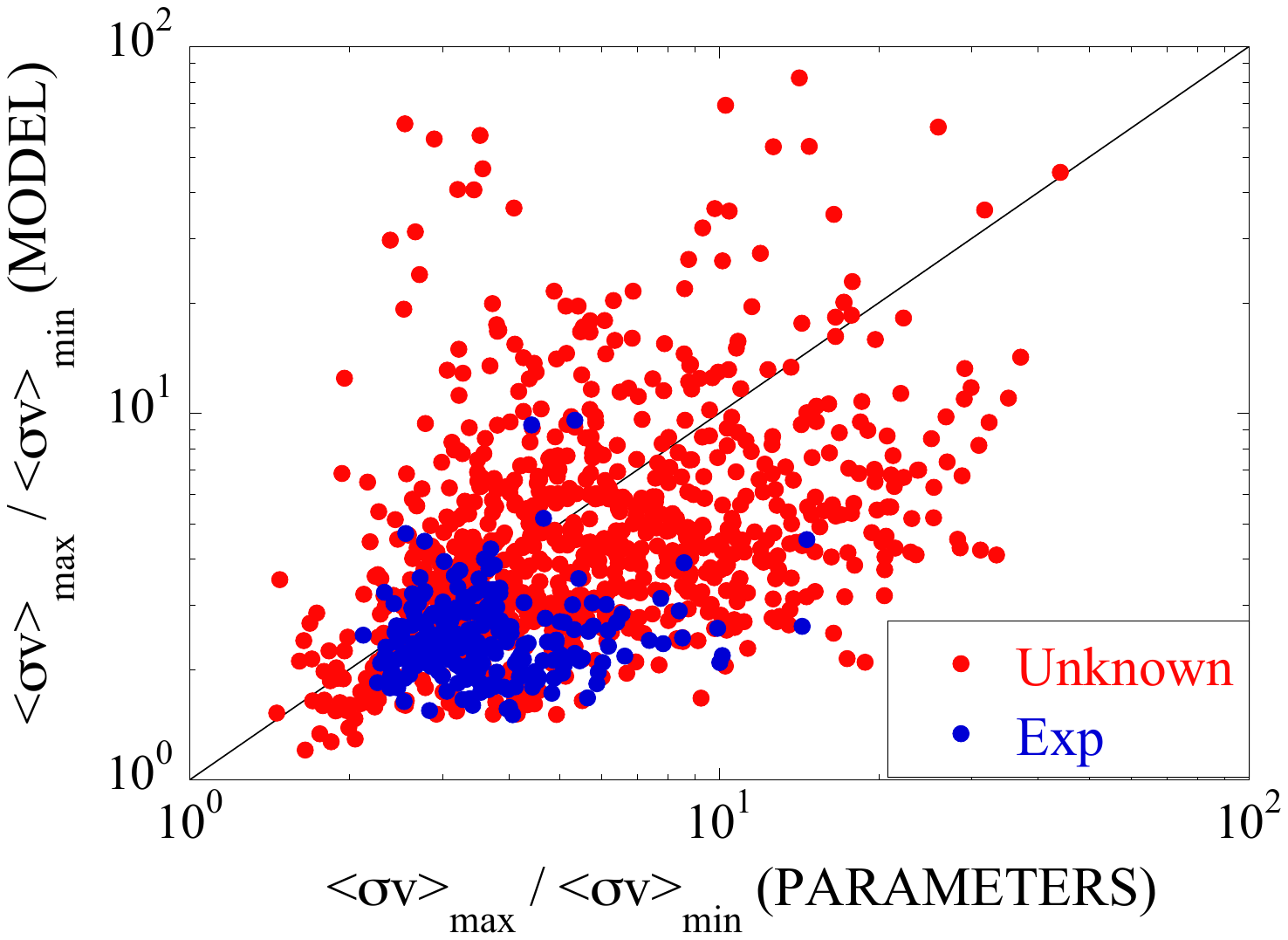}
    \caption{Correlated model uncertainties {represented by the ratio of the maximum to minimum \nga MACS} (Fig.~\ref{fig:parameters_uncertainties_Sn}a) as a function of non-correlated parameter uncertainties of set B, (Fig.~\ref{fig:parameters_uncertainties_Sn}b). In blue the 239 experimentally known \nga reactions and in red the 868 theoretical ones. All rates are estimated at $T=2.5 \times 10^8$~K.}
    \label{fig:parameters_uncertainties_vs_model_uncertainties}
\end{figure}

Figure \ref{fig:parameters_uncertainties_vs_model_uncertainties} illustrates the correlated model uncertainties against the non-correlated parameter uncertainties, as extracted from Fig.~\ref{fig:parameters_uncertainties_Sn}. If both uncertainties were correlated, we would expect each reaction to lie along the 1:1 ratio line depicted in black. Here we can see a broad dispersion, where reactions with a high model uncertainty have a low parameter one, and conversely, reactions that are described similarly by different nuclear models and showing a high parameter uncertainty. This underlines the need to take into account both the correlated model and the non-correlated parameter uncertainties.
It also emphasizes the fact that maximum deviations estimated from model variations are not suited to describe uncorrelated nuclear uncertainties, as sometime considered \citep{mckay20}.

\begin{figure*}
    \centering
    \includegraphics[width=0.48\textwidth]{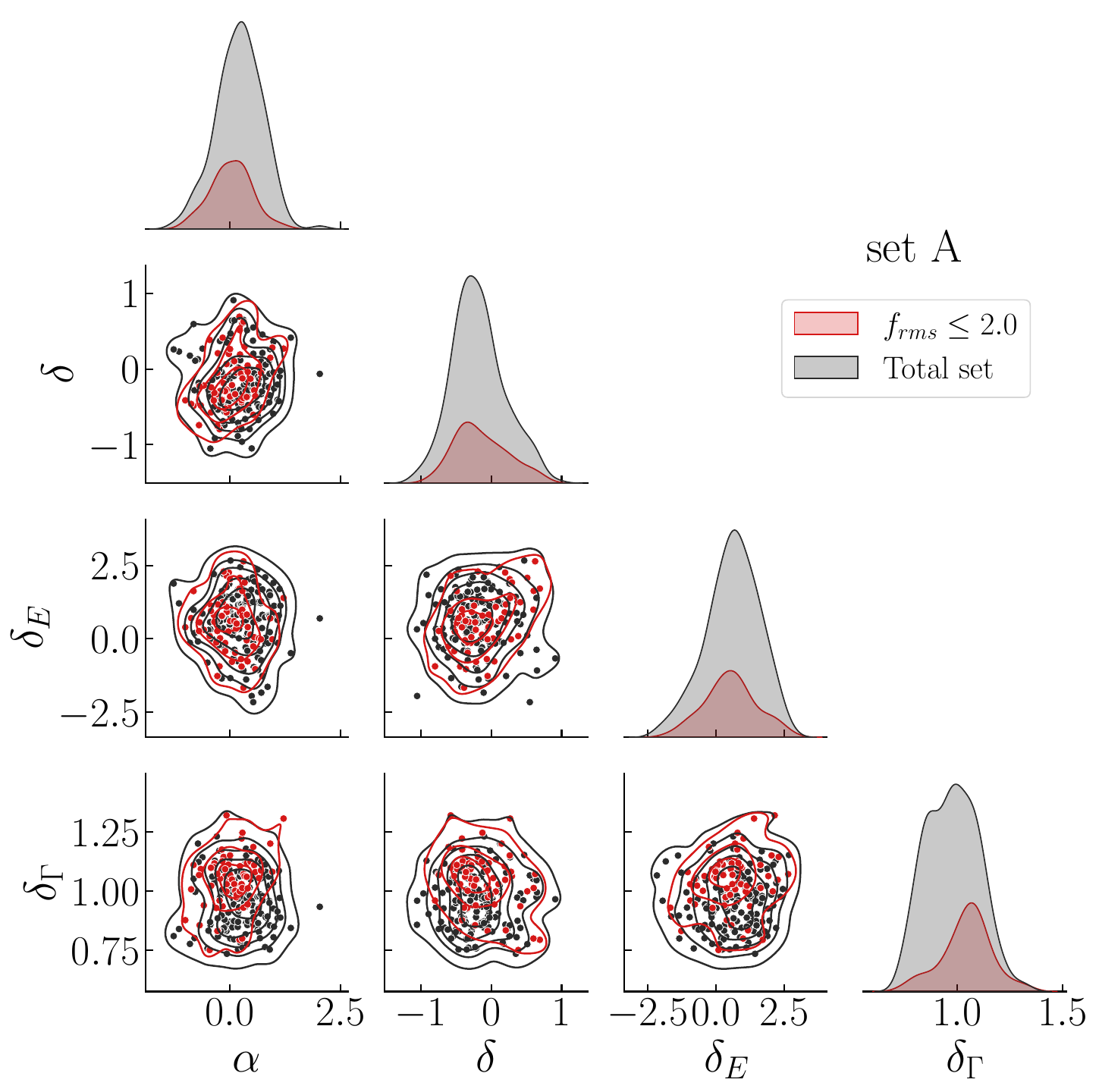}
    \includegraphics[width=0.48\textwidth]{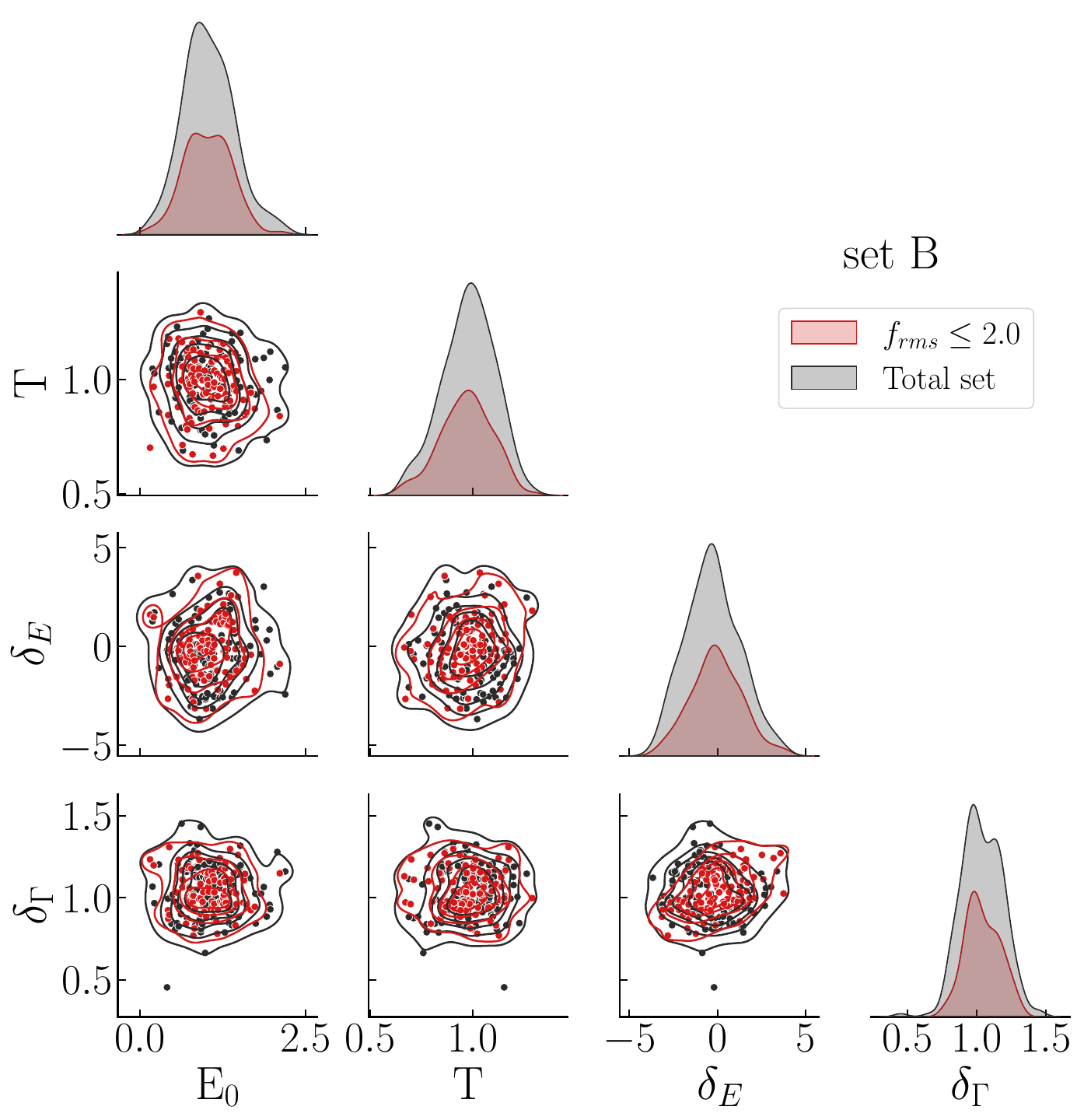}
   \caption{Correlation plots for the 4 parameters in set A (left) and set B (right) using the total set ($N_{\rm comb}$=200; black dots) or the selected subset with $f_\rms\leq2.0$ ($N_{\rm comb}$=61 for set A and $N_{\rm comb}$=97 for set B; red dots). Each contour corresponds to iso-proportions of the density, with the first contour corresponding to 20\% and incrementing by 20\%. The diagonal shows the (normalized) marginal distribution of each parameter. }
   \label{fig:pairplot_correlation}
\end{figure*}

Figure \ref{fig:pairplot_correlation} shows the correlation between the 4 different parameters used in the selected combinations resulting to $f_\rms \leq$ 2.0 for both sets A and B. Set A parameters $\alpha$ and $\delta$ ($E_0$ and $T$ are the equivalent in set B) are local variations of parameters impacting the NLD, while $\delta_E$ and $\delta_{\Gamma}$ impact the PSF. The diagonal shows the marginal distribution of each parameter. As expected from the random multivariate distribution, the parameters are mostly normally distributed around their nominal value. This shows that we explore consistently the parameter space resulting in combinations with $f_\rms \leq 2$. 
For set A, we observe a correlation between ($\alpha$, $\delta$) and $(\delta_E$, $\delta_{\Gamma}$), as expected since the NLD (hence the MACS) increases for increasing $\alpha$ or decreasing $\delta$ values and the E1 PSF (hence the MACS) also increases for increasing $\delta_{\Gamma}$ or decreasing $\delta_E$ values. For the same reason, an anti-correlation between $\delta$ and $\delta_E$ also appears. For set B, there is also a clear correlation between $\delta_E$ and $\delta_{\Gamma}$. Other correlations are however less pronounced for this set.

\section{Impact on the i-process nucleosynthesis}

To explore the impact of the correlated model and non-correlated parameter uncertainties, we compute a large set of multi-zone stellar evolution models during the early AGB phase of a low-mass low-metallicity star using the STAREVOL code \citep{siess00,siess2008}, as described below. Note that in 1D stellar models, the star is divided into spherical zones (or shells) in which the stellar structure equations and nucleosynthesis are solved. The zones can interact with each other through mixing processes such as convection. In our models, during the PIE, the star is divided into $\sim 3500$ zones, of which around $500$ are dedicated to the convective thermal pulse where the i-process takes place.

\subsection{The proton ingestion episode and the dilution procedure}
\label{sec:astromod}
The 1~$M_{\odot}$, [Fe/H]~$=-2.5$ AGB model considered here was already extensively discussed in \cite{choplin21, goriely21, Choplin2022letter, choplin22}. We only recall here a few important evolutionary aspects and explain the dilution procedure we devised to save computational time.

During the early TP-AGB phase, protons are engulfed by the convective thermal pulse and burn by the $^{12}$C($p,\gamma$)$^{13}$N reaction while being transported down. After the $\beta^+$-decay of $^{13}$N into $^{13}$C in a timescale of $10$~min, the $^{13}$C($\alpha,n$)$^{16}$O reaction is activated at the bottom of the convective pulse at a temperature of about 250~MK and leads to neutron densities of $\sim 10^{15}$~cm$^{-3}$. Shortly after the neutron density peak, the convective pulse splits \citep[][Sect 3.5 for more details on the split]{choplin22}. After the split, the upper part of the convective pulse grows in mass and engulfs additional protons. However, the temperature at the bottom of the upper part of the pulse is now too low to activate the $^{13}$C($\alpha,n$)$^{16}$O reaction efficiently. So, no substantial modification of the abundances of heavy elements is found after the split. However, because H-burning is still operating, species involved in the CNO cycle will keep evolving. The upper part of the pulse eventually merges with the convective envelope, leading to the enrichment of the surface in i-process products (and other C and N isotopes). 

The PIE in our reference AGB model was computed for each set of \nga rates.
To save computational time, the models are stopped {0.7~yr} after the split, before {the elements are brought to the surface}. Nevertheless, since the i-process nucleosynthesis is restricted to the evolution before the split, the surface abundances can still be estimated accurately thanks to the dilution procedure explained below. 

In our reference model, {just} after the split, we {calculate} the mean abundance of each element ($j$) in the upper part of the convective pulse (which {did not have time to homogenize yet}) as 
\begin{equation}
\bar{X}^\mathrm{pulse}_j = \frac{1}{M_2-M_1} \int_{\rm M_1}^{\rm M_2} \, X_j \, dm
\end{equation}
where $M_1$ and $M_2$ are the mass coordinate boundaries of the upper convective pulse (with $M_1$ very close to the mass coordinate of the split). This material is then diluted into the envelope (which is still disconnected from the pulse at this time), leading to the approximate surface mass fraction 
\begin{equation}
X^{\rm dil}_j = \bar{X}^{\rm pulse}_j (1-f_{\rm dil}) + {X}^{\rm env}_j f_{\rm dil}.
\end{equation}
{where ${X}^{\rm env}_j$ is the (homogeneous) envelope mass fraction of element $j$.}
The dilution parameter $f_{\rm dil}$ is adjusted {on our reference model} so as to minimize the difference between the {exact} final surface abundances $X^{\rm surf}$ and {its estimate} $X^{\rm dil}$ (all abundances are considered after $\beta$- and $\alpha$-decays). 
{For our AGB model}, $f_{\rm dil}$ {is equal} to $0.9243$, which leads to a deviation $| X^{\rm surf} - X^{\rm dil}| / X^{\rm surf} < 0.04$ {for all elements, except for Li (deviation of 0.99), C (deviation of 0.09) and N (deviation of 0.76)}.
All the stellar models considered in this work have almost exactly the same structures and follow the same evolutionary pathway. Hence, this calibration of $f_{\rm dil}$ can be safely used for all of them, {as shown by our tests.}

\begin{figure*}
    \centering
    \includegraphics[width=0.85\textwidth]{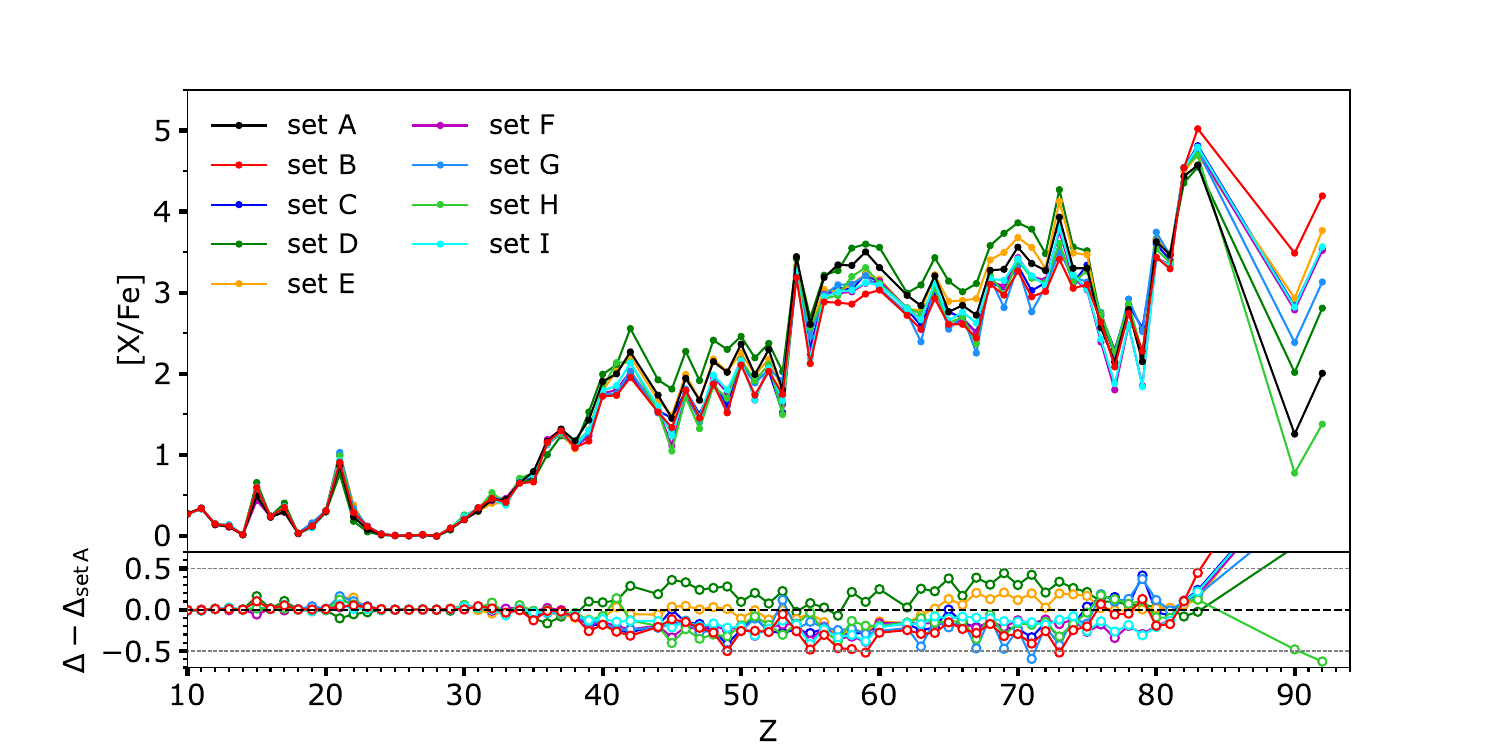}
    \caption{ 
    Surface [X/Fe] abundance ratios of AGB models for the 9 different nuclear models considered. The bottom plot shows the differences between the various models and set A. 
    }
    \label{fig:xfe_syst}
\end{figure*}
\subsection{Propagating model uncertainties}
\label{sec:propmod}

\begin{figure}
    \centering
    \includegraphics[trim={1.0cm 1.0cm 0.4cm 0.7cm},width=0.46\textwidth]{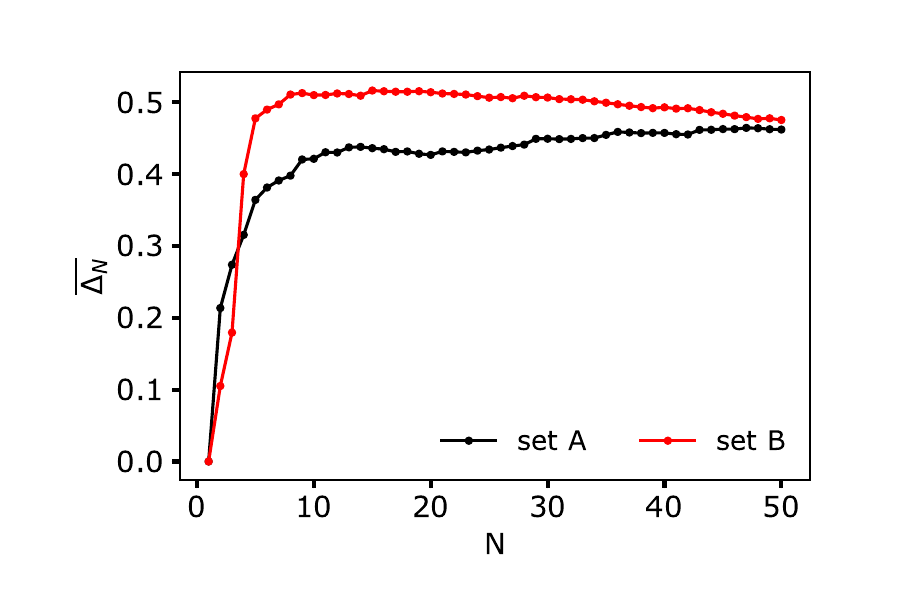}
\caption{Evolution of the averaged uncertainties $\overline{\Delta_N}$ (as defined in Eq.~\ref{eq:del}) as a function of the number of simulations $N$.} 
\label{sec:proppar}
    \label{fig:deldel_xfe}
\end{figure}

To propagate the correlated model uncertainties, we use here the same methodology, discussed in \citet{goriely21}, {consisting in PIE stellar} calculations (as described in Sect.~\ref{sec:astromod}) for the 9 different nuclear models introduced in Sect.~\ref{sec:mod}. Note that in comparison with our previous work \citep{goriely21}, alternative combinations of PSFs and NLDs are adopted here but uncertainties associated with the contribution of the direct capture mechanism are not taken into account. Sets A, D, E, F and I essentially test the impact of PSF models using the same NLDs, while sets A, C, G and H illustrate the impact of the NLD models using the same PSF.
Figure \ref{fig:xfe_syst} shows the resulting surface abundances [X/Fe] and the corresponding deviations stemming from the correlated model uncertainties for the 9 different nuclear models, including sets A and B for which parameter uncertainties are also explored in more details in Sect.~\ref{sec:proppar}. The lower panel shows the difference between the various models and set A.  Overall, the correlated nuclear model uncertainties impact the chemical abundances of $Z>40$ elements by typically $0.5 - 1$ dex with the exception of Th ($Z=90$) and U ($Z=92$) {for which the effects are stronger}. The potential production of Th and U is highly sensitive to the model uncertainties and can lead to orders-of-magnitudes differences,  {but variations of [Th/U] remains constrained to values between $-0.84$ to $-0.61$}, as discussed in \citet{Choplin2022letter}. 
Some nuclear models can lead to quite different surface abundance predictions, especially for important tracers such as Ba ($Z=56$), La ($Z=57$) and Eu ($Z=63$). However, since the model uncertainties are correlated, abundance ratios may be significantly less affected. In particular, the [Ba/Eu] ratio varies from 0.12 to 0.58 and the [Ba/La] ratio from -0.15 to 0.02.
Some nuclear models also give rise to systematically different results, in particular set D with its relatively slow rates obtained with the low GLO PSFs tends to overproduce elements between $Z=40$ and $Z=75$ in comparison with other models.


\subsection{Propagating parameter uncertainties}

\begin{figure*}
    \centering
    \includegraphics[width=0.85\textwidth]{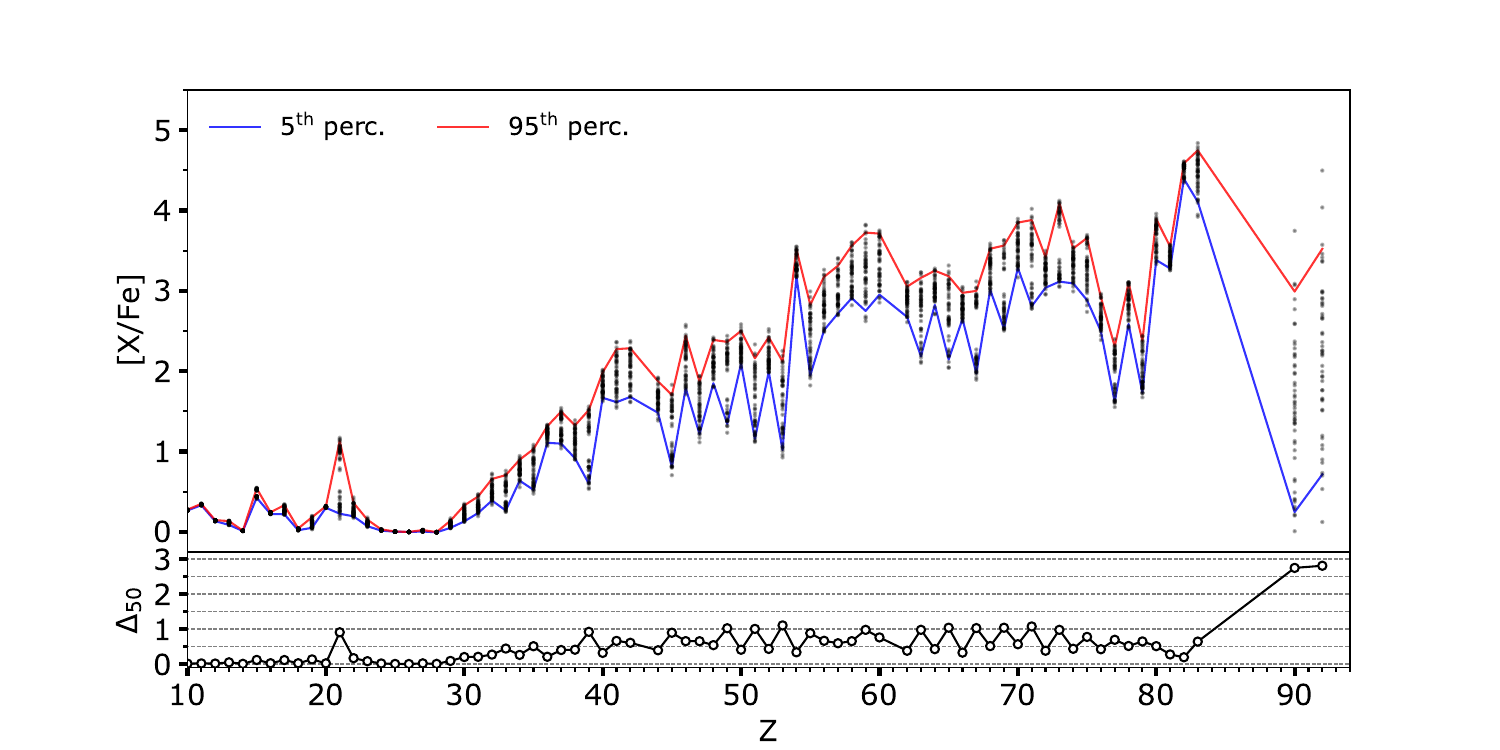}
    \includegraphics[width=0.85\textwidth]{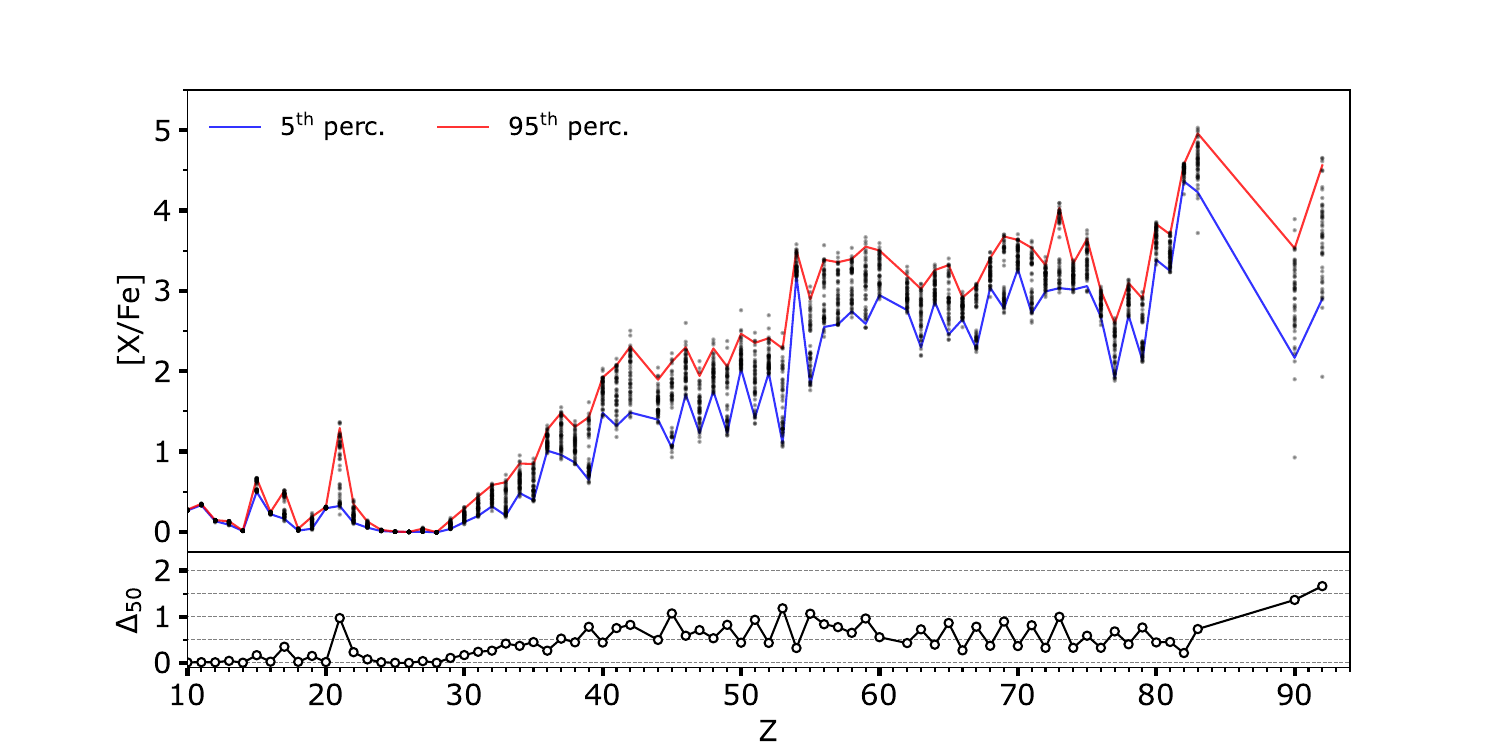}
    \caption{Scatter plot of the surface [X/Fe] abundance ratios of AGB models for the nuclear set A (top) and set B (bottom) with $f_\rms$ $\leq 2.0$. The blue (red) line shows the $5^{\rm th}$ ($95^{\rm th}$) percentile (Eq.~\ref{eq:ddel}). The bottom subplot in each panel shows $\Delta_{50}$, which is the difference between the $95^{\rm th}$ and $5^{\rm th}$ percentile (Eq.~\ref{eq:ddel}) for the $N=50$ simulations.}
    \label{fig:xfe_perc_frms}
\end{figure*}

\begin{figure*}
    \centering
    \includegraphics[width=0.85\textwidth]{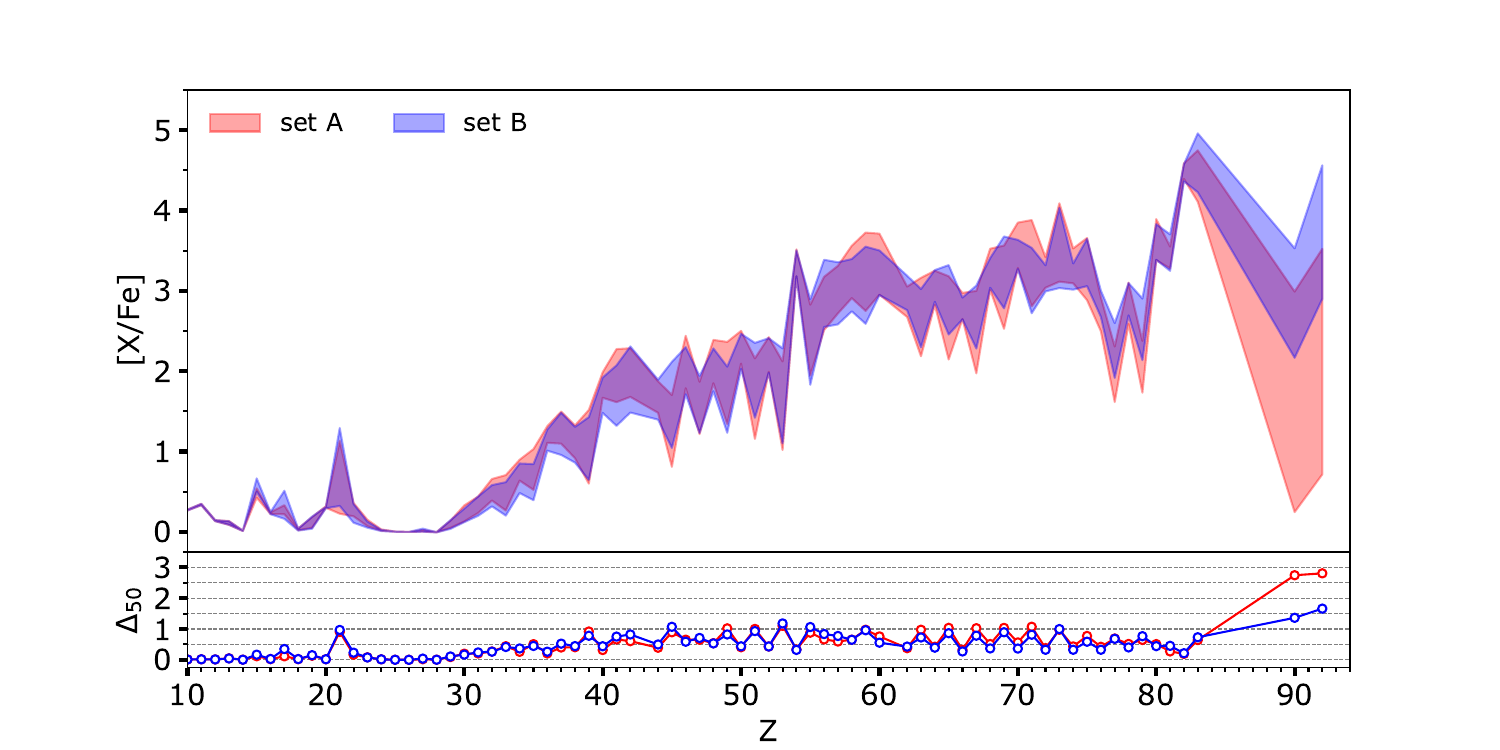}
    \caption{Comparison between the surface abundances [X/Fe] and their uncertainties obtained with nuclear sets A and B, for $f_\rms\leq2.0$ case. The red (blue) region shows the extent of the [X/Fe] region between the 5th and 95th percentile for set A (set B). The purple area shows the intersection between the two models.}
    \label{fig:xfe_comp_mod}
\end{figure*}

To propagate the uncorrelated parameter uncertainties, a large number $N$ of sets containing randomly chosen minimum and maximum rates for each of the 868 \nga reactions is produced. These are based on the maximum and minimum rates obtained in Sect.~\ref{sect:forward_MC} from the BFMC method. We can choose randomly between these rates due to the non-correlated nature of these uncertainties. These random sets of rates are then used in stellar evolution models to assess the impact of the parameter uncertainties on the final surface abundances in our 1~$M_{\odot}$  [Fe/H]~$=-2.5$ AGB star. {Despite the fact that parameter uncertainties are uncorrelated, the resulting surface abundances may remain correlated for a given stellar simulation since a given rate may influence more than one isotopic abundance.}

One of the difficulties with randomly produced sets of rates is to ensure that enough draws have been made to be representative of the full range of possible outcomes. We computed an increasing number of stellar evolution models until convergence of the upper and lower limits of the surface abundances is reached.
More specifically, the convergence is evaluated by the quantity
\begin{equation}
\overline{\Delta_N} = \frac{1}{N_Z}\sum_{j=1}^{N_Z} \Delta_{N}
\label{eq:del}
\end{equation}
where $N_Z$ is the total number of elements and
\begin{equation}
\Delta_N = p_{N}^{95} (\mathrm{[X/Fe]}) - p_{N}^{5} (\mathrm{[X/Fe]})
\label{eq:ddel}
\end{equation}
where $p^{95}$ and $p^{5}$ refer to the 95$^{th}$ and 5$^{th}$ percentiles, respectively, of the [X/Fe] distributions
and $N$ to the number of AGB i-process simulations considered (up to 50) to estimate these percentiles. We choose to select uncertainties values from the 5$^{\rm th}$ and 95$^{\rm th}$ percentiles to take into account potential numerical artifacts that could lead to spurious abundances over- or under-estimates. $\Delta_{50}$ corresponds to the difference between the percentiles when considering 50 different simulations (hence 50 different sets of randomly chosen rates among their maximum and minimum values).

Figure~\ref{fig:deldel_xfe} shows the convergence of the averaged uncertainties $\overline{\Delta_N}$ as a function of the number $N$ of simulations. For small values of $N$, a rapid increase is expected due to the random nature of the draws, leading to large variations in the uncertainties. For  more than typically $N=30$ simulations, a plateau is found and the global abundance uncertainty does not evolve anymore when compared to the value for $N=50$ simulations\footnote{Note that the percentile-dependent $\overline{\Delta_N}$ can be non-monotonically increasing due to the appearence of outliers. }. In other words, adding an extra 20 simulations does not give rise to average changes of more than 0.05 dex. Hence, we consider that we have convincingly converged to the total propagation of parameter uncertainties by computing $N=50$ stellar simulations with 50 different nuclear sets.  

\subsubsection{Impact on surface abundances}
\label{sec:xfe-syst}

Figure~\ref{fig:xfe_perc_frms} shows the surface [X/Fe] abundances resulting from the $N=50$ simulations computed for the nuclear set A (top panel) and set B (bottom panel). For a given element, each black dot corresponds to the final abundance of one of the 50 simulations. 
The uncertainties on the i-process abundances are of the order of 0.5 to 1.0 dex on average for all the nuclei with $Z\geq 40$. Interestingly, we can discern a clear pattern of higher uncertainties for the odd-$Z$  elements. This comes from the fact that odd-$Z$ elements have usually only one stable isotope, so that these isotopes are highly sensitive to the competition between the \nga reaction and the $\beta$-decay of the even $Z-1$ isotopes. An eye-catching example is the case of Sc ($Z=21$) where we can clearly see two groups of abundances separated by almost 1 dex in set A. It clearly underlines the sensitivity to a rate being maximum or minimum and acting directly on the final Sc abundances. We show in Sect. \ref{Sect:important_reaction} that, in the Sc case, this is due to the $^{45}$Ca\nga reaction.

{Since the surface abundances may remain correlated, the uncertainies affecting abundance ratios should be estimated from the maximum and minimum abundance ratios among the 50 different simulations.}
{The range of possible [Ba/Eu] ratios for the nuclear set A (set B) is $-0.55 < \mathrm{[Ba/Eu]} < 0.85$ ($-0.46 < \mathrm{[Ba/Eu]} < 0.79$), {\it i.e.} an uncertainty of 1.35 dex (1.25 dex). For the [Ba/La] ratio, the ranges become $-0.67 < \mathrm{[Ba/La]} < 0.31$ for set A (uncertainty of 0.98 dex) and $-0.75 < \mathrm{[Ba/La]} < 0.63$ for set B (uncertainty of 1.38 dex).
The [Th/U] ratio can vary from $-1.31$ to $-0.11$ for set A (uncertainty of 1.20 dex) and from $-1.22$ to $-0.14$ for set B (uncertainty of 1.08 dex).}
The abundance ratios are consequently significantly more affected by parameter than model uncertainties (see Sect.~\ref{sec:propmod}). It is consequently of prime importance to decrease the parameter uncertainties, especially on tracer elements, like La, Ba or Eu.

Sets A and B lead to similar uncertainty patterns, as confirmed in Fig.~\ref{fig:xfe_comp_mod}, which compare, for both sets, the surface abundance predictions [X/Fe] and their respective uncertainties. The shaded red and blue zones correspond to set A and set B, respectively, the limits being given by the 5$^{\rm th}$ and 95$^{\rm th}$ percentiles. We can see that the amplitude of the uncertainties from both nuclear sets are comparable for almost all nuclei. A clear difference of more than 1 dex affects the production of Th and U. This underlines the uncertainties still impacting the nuclear physics predictions of sub-actinides and actinides. Otherwise, both nuclear sets lead to fairly close values of the surface abundances. 

\begin{figure*}
    \centering
    \includegraphics[width=0.85\textwidth]{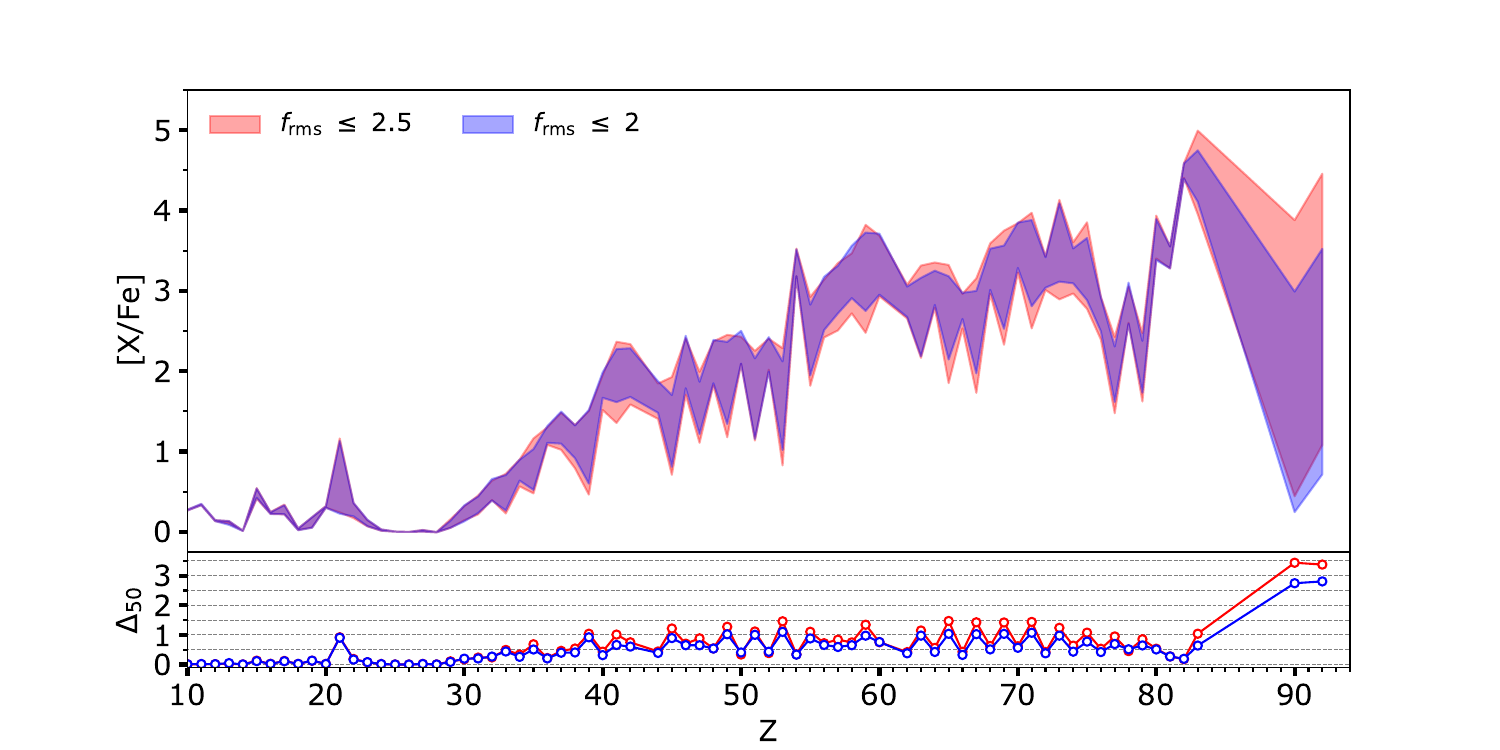}
    \caption{Comparison between the surface abundances [X/Fe] and their uncertainties obtained with the non-correlated parameter uncertainties of nuclear set A with $f_\rms \leq$ 2.0 (blue) and $f_\rms \leq$ 2.5 (red).}
    \label{fig:xfe_comp_frms}
\end{figure*}


\subsubsection{Impact of the $\chi^2$ threshold value}

We additionally explored the impact of the BFMC $\chi^2$ criterion, as described in Sect.~\ref{Sect:backward_MC} ($\chi^2 \leq \chi_{\rm crit}^2$) by increasing the value of $\chi_{\rm crit}^2$ to an upper $f_\rms$ deviation of 2.5 instead of 2.0. To do that, we extend the {sets} of 4-parameter combinations to a $f_\rms~\leq~$2.5 ($N_{\rm comb}$=112 for set A and $N_{\rm comb}$=161 for set B) to estimate the 868 theoretical MACS. With the corresponding randomly chosen upper and lower limits of these rates in each of the sets A and B, we computed once again 50 stellar evolution models. The extended parameter combinations obviously allow for higher maximum and lower minimum rates, hence a higher impact on the surface abundances can be expected. Figure \ref{fig:xfe_comp_frms} compares the surface [X/Fe] abundances and their parameter uncertainties (5$^{\rm th}$ and 95$^{\rm th}$ percentiles) computed with rate combinations restricted to $f_\rms \leq 2.5$ and  $f_\rms \leq 2.0$, both for the nuclear set A. 
Interestingly, the uncertainties increase mostly for $Z \geq 40$ odd-$Z$ elements from $\sim 1.0$ dex for an $f_\rms \leq 2.0$ to $\sim 1.5$ dex for $f_\rms \leq 2.5$. However, the even-$Z$ elements retain the same uncertainties of about 0.5 dex as for $f_\rms \leq$ 2.0. Similar results are found when adopting set B.    


\section{Important reactions and their impact}

\subsection{Determining the most impacting reactions}
\label{Sect:important_reaction}

We explore the potential impact of given reactions on the i-process nucleosynthesis by taking advantage of the method we use to assess the parameters uncertainties. Indeed, the abundance distribution obtained in our stellar simulations for each element can be used to separate random from correlated effects. More specifically, for each nuclei we can sort the abundances {from our 50 simulations} and obtain the corresponding sorted list of maximum/minimum rate distributions for each of the 868 \nga reactions. If the abundance of a given nucleus is not sensitive to any reaction, the corresponding sorted distribution of rates should be a random distribution of maximum and minimum values. However, if this sorted distribution {leads}, in the extreme case, to 25 consecutive maxima followed by 25 consecutive minima (a very unlikely random draw), it suggests a correlation between the value of a given reaction rate and the surface abundance of the concerned nucleus. To quantify the likeliness of such a draw, we use the properties of what is in fact here a geometrical distribution. The probability to draw consecutively a same rate (only maxima or only minima) is defined as:
\begin{equation}
P(k)=(1-p)^k p
\label{eq:rand_like}
\end{equation}
with $k$ the number of consecutive equal rates, and $p$ the probability of the trial, here $p=0.5$ as we draw randomly between maximum and minimum rates. In the case of 25 consecutive draws, the probability to obtain such a distribution is then $P(25)\simeq 10^{-8}$, meaning there is 1 chance over 100 millions to draw such a distribution, a more than highly unlikely draw for a $N=50$ distribution. Using this ``random likelines'' criterion, we check for each nucleus in our network if their sorted abundance value is correlated to their corresponding sorted rate distribution. 

Tables \ref{Table:main} and \ref{Table:tracer} present the different elements for which the surface abundance is directly impacted by a given reaction. 
While Table \ref{Table:main} lists the reactions, sorted by decreasing values of the surface abundance uncertainty for set A, Table \ref{Table:tracer} shows a subset of this table, focusing on the elements that are observable in the atmosphere of CEMP stars, and that can be used as tracers of the i-process nucleosynthesis. The adopted selection criterion assumes that at least two consecutive sequences of draws with both a random likeliness of $P\leq10^{-3}$ {(i.e $k \ge 8$)} are found in the sorted distribution. 

\begin{figure*}[h!]
    \centering

    \includegraphics[height=0.27\textheight]{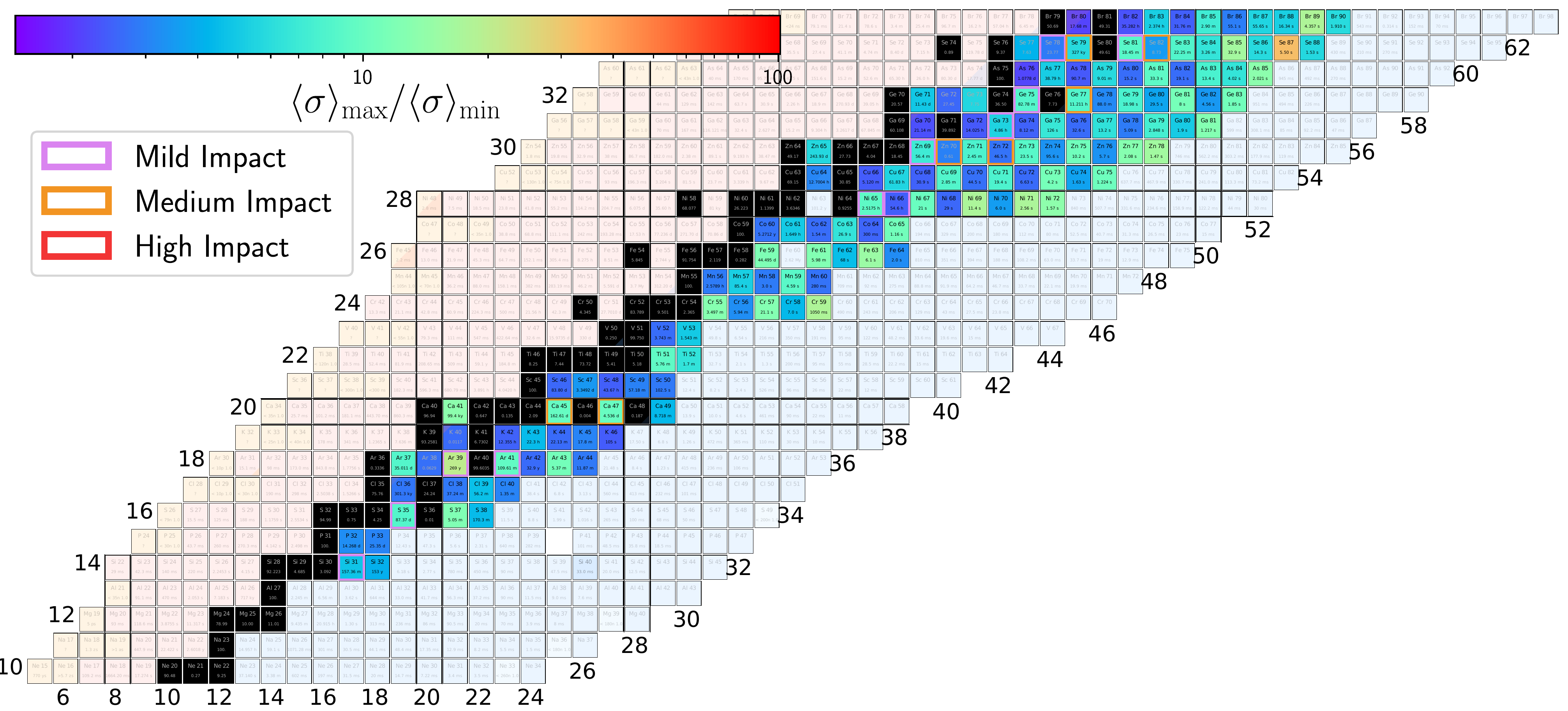}\vspace{-0.8cm}
    \includegraphics[height=0.27\textheight]{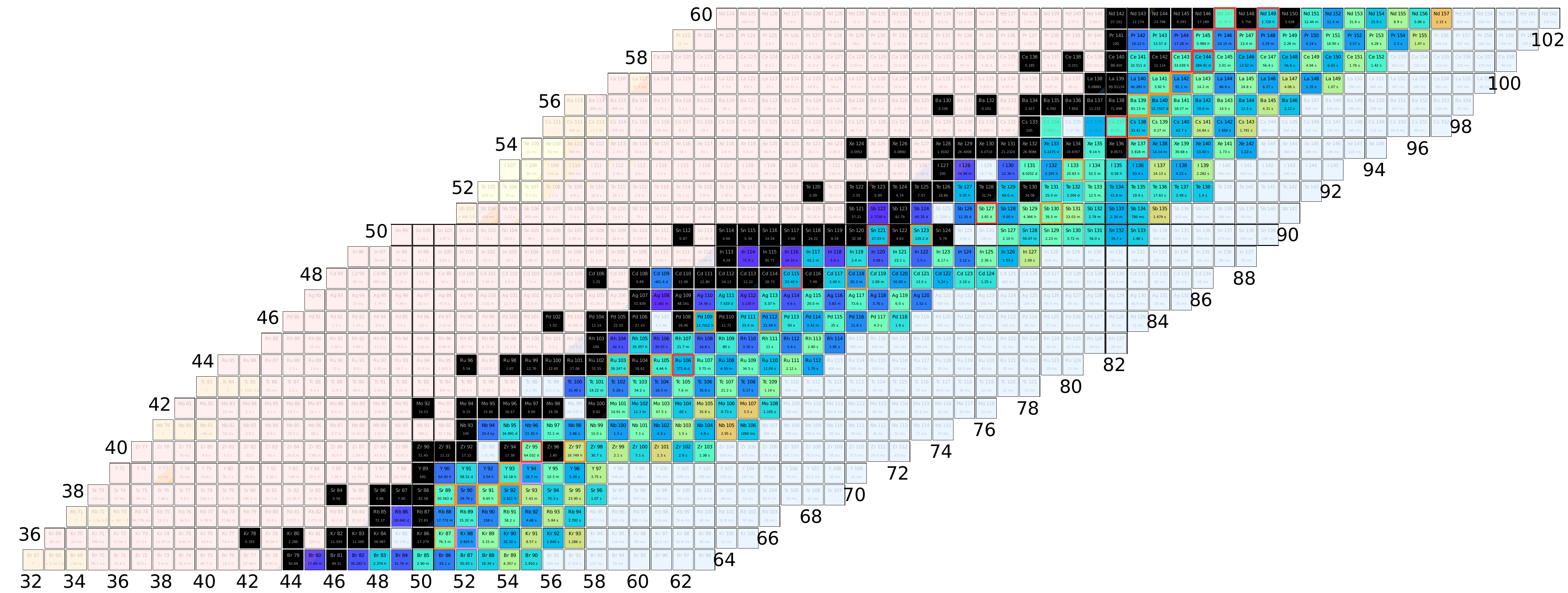}\vspace{-0.55cm}
    \includegraphics[height=0.25\textheight]{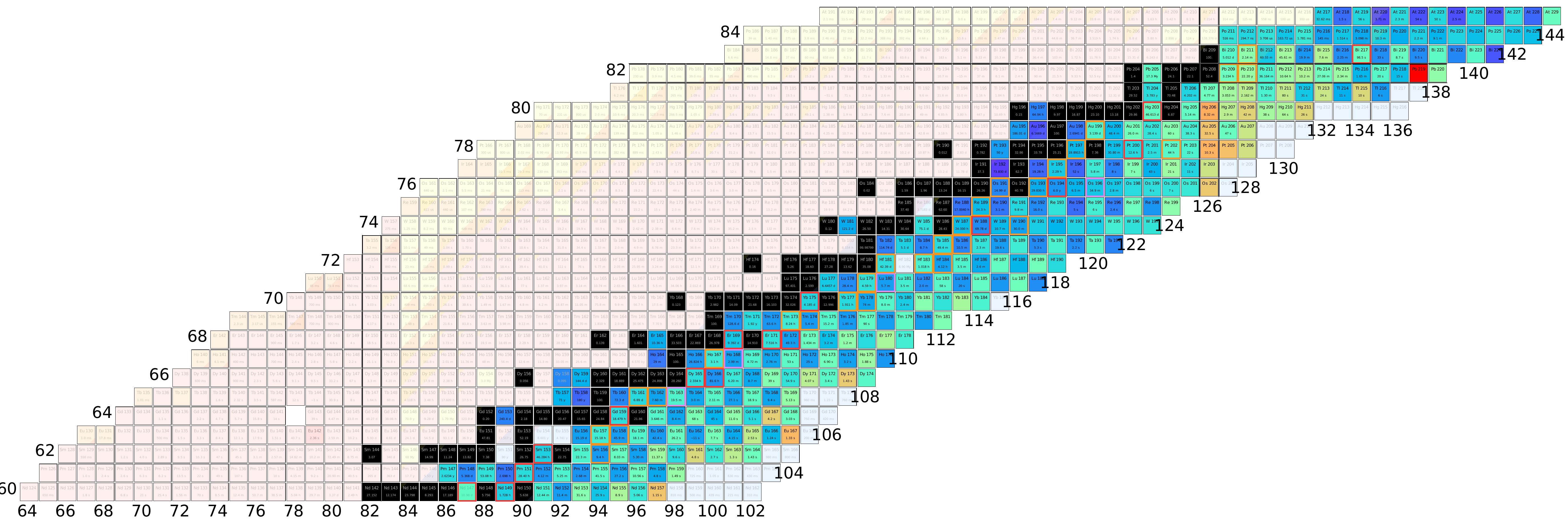}\vspace{-1.5cm}
    \includegraphics[height=0.18\textheight]{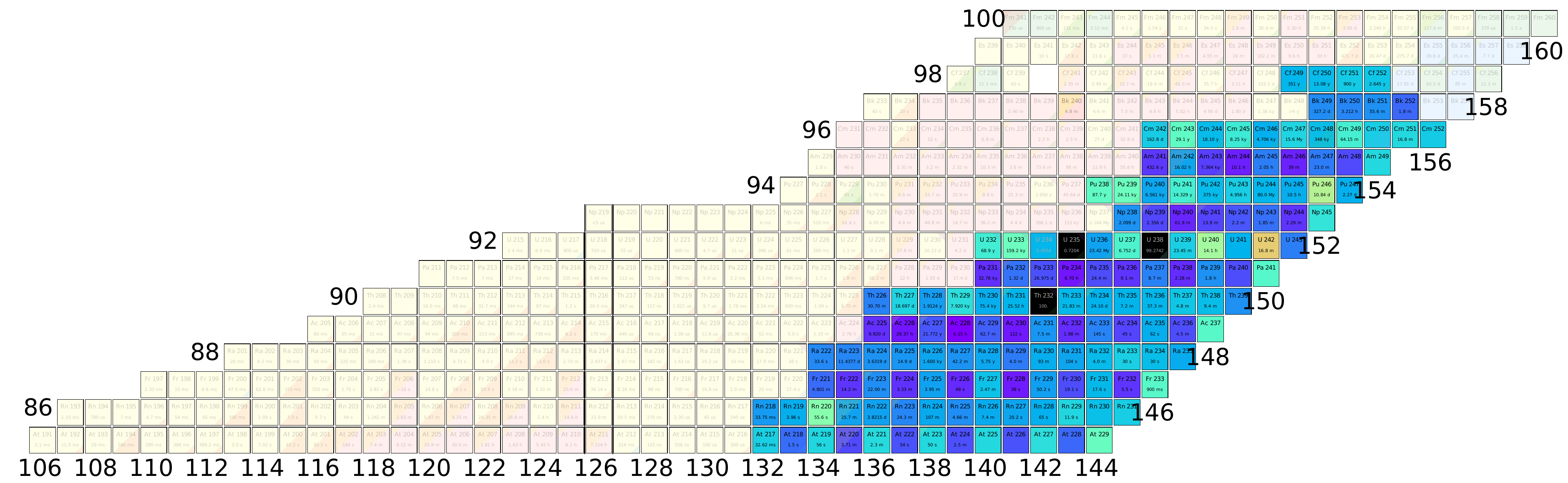}\vspace{0.5cm}
\caption{Chart of the nuclides highlighting the 868 experimentally unknown \nga reactions relevant to the i-process nucleosynthesis. The reaction are color-coded by their non-correlated parameter uncertainty (the ratio between maximum and minimum rates) for the nuclear set B. Some of the reaction targets are highlighted by a square contour of pink/orange/red color. These correspond to reaction targets, as given in Table~\ref{Table:main} (Sect. \ref{Sect:important_reaction}) for which the parameter uncertainties have a mild/medium/high impact, respectively, on one or multiple isotopic abundances.}
    \label{fig:NZ_plan}
\end{figure*}

Figure \ref{fig:NZ_plan} illustrates the chart of nuclides with emphasis on the 868 experimentally unknown \nga reactions relevant to the i-process nucleosynthesis. The reactions are color-coded by their non-correlated parameter uncertainty (the ratio between maximum and minimum rates) for the nuclear set B obtained from the BFMC method in Sect. \ref{sect:forward_MC}.{The nuclei} highlighted by a square pink/orange/red color contour {identify elements for which the \nga reaction} have a mild/medium/high impact $x$ (defined as 0.1 dex $\leq x <$ 0.5 dex / 0.5 dex $\leq x <$ 1.0 dex / $ x \geq$ 1.0 dex,  respectively) on one or multiple isotopic abundances and have a maximum isotopic fraction larger than 15\%. Most of the highly uncertain \nga reaction rates are away from the stability zone. However, some of the rates close to the valley of $\beta$-stability can also be quite uncertain, leading to a large impact on abundances, especially for targets with an even-$Z$ number. Indeed, these reactions compete with the $\beta$-decay to an odd-$Z$ number that usually have only one stable isotope. As a consequence, the rate has a significant influence on whether or not the flux will feed the stable isotope provided the \nga rate is lower than the $\beta$-decay rate, or if the flux will pursue further away from this isobar if the \nga rate is higher.

\subsection{Impact of reducing the nuclear uncertainties}
\label{Sect:direct_impact_reactions}

Now that the most impacting reactions have been determined, it remains to estimate their quantitative effect on the surface abundances. We could estimate the impact of each reaction entering Tables \ref{Table:main} and \ref{Table:tracer} by running an two additional  simulations with the upper and lower limits of that specific rate only. This procedure is of course costly in terms of computing, so that only some illustrative examples are given below. These concern the recently constrained $^{139}$Ba($n,\gamma$)$^{140}$Ba rate directly affecting the s- and i-tracer La, the $^{153}$Sm($n,\gamma$)$^{154}$Sm acting on the production of the r- and i-tracer Eu and $^{217}$Bi($n,\gamma$)$^{218}$Bi conditioning the production of Th and U.

\subsubsection{The $^{139}$Ba($n,\gamma$)$^{140}$Ba reaction}

\begin{figure}
    \centering
    \includegraphics[width=0.42\textwidth]{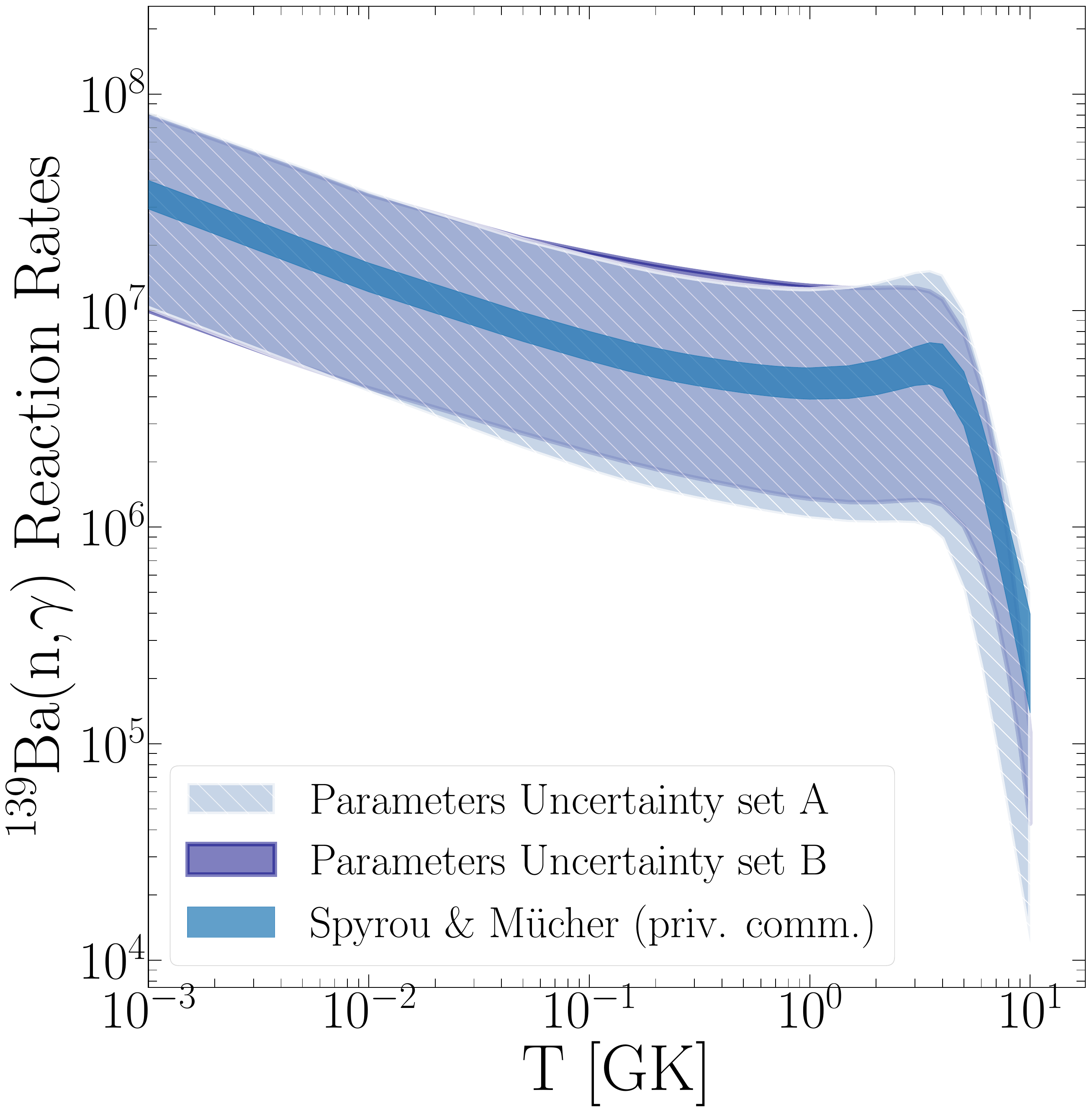}
    \caption{Theoretical parameter uncertainties vs newly measured $^{139}$Ba\nga rate (M\"ucher \& Spyrou, priv. comm).}
    \label{fig:Ba139_rates}
\end{figure}

The $^{139}$Ba($n,\gamma$)$^{140}$Ba reaction rate has recently been constrained experimentally by M\"ucher \& Spyrou (priv. comm). It corresponds to one of the important reactions found in Sect. \ref{Sect:important_reaction} impacting the production of $^{139}$La, a relevant observable tracer (see Table~\ref{Table:tracer}).
Figure \ref{fig:Ba139_rates} shows the parameter uncertainties on the $^{139}$Ba\nga rate for sets A and B\footnote{Note that the impact of the direct capture mechanism on this specific reaction has been neglected since it is estimated {not to exceed} 10\%, see in particular Fig. 4 of \citet{goriely21}.}, as well as the new experimental constraint of M\"ucher \& Spyrou (priv. comm). We can see a clear reduction of the theoretical uncertainties, reaching 85\% around the temperature of 250MK at which the i-process nucleosynthesis takes place in AGB stars.
The $^{139}$La abundance uncertainties obtained for set B amounts to 0.77 dex {and is unlikely due to the sole variations of the $^{139}$Ba\nga rate. Potentially complex combinations of different maximum and minimum rates around La and Ba isotopes will also contribute.} To quantify this impact, we run two stellar simulations, this time only varying the $^{139}$Ba\nga rate to evaluate its impact on the $^{139}$La production. 
For this $^{139}$Ba\nga analysis, we use the geometric mean between the maximum and minimum rates for all the 867 unknown \nga rates of set B and keep them fixed while running our two simulations with  minimum and the maximum values of the $^{139}$Ba\nga rate for set B. Doing so, the parameter uncertainty associated to the $^{139}$Ba\nga rate is found to affect the $^{139}$La production by 0.4 dex out of the 0.77 dex of total uncertainty (obtained when allowing all rates to be changed within their lower and upper limits in set B). 
The impact on the La production resulting from the reduction of uncertainties thanks to the recent measurement of M\"ucher \& Spyrou can be estimated in the same way. New stellar evolution simulations, still using the geometric mean rates for the 867 theoretical ones and the upper and lower limits of the experimentally constrained rate now lead to an uncertainty of less than 0.06 dex on the La production to be compared to the above-mentioned 0.4 dex uncertainty. This test shows how relevant this $^{139}$Ba($n,\gamma$)$^{140}$Ba reaction is for an accurate prediction of the La synthesis by the i-process.


\begin{figure*}
    \centering
    \includegraphics[width=0.85\textwidth]{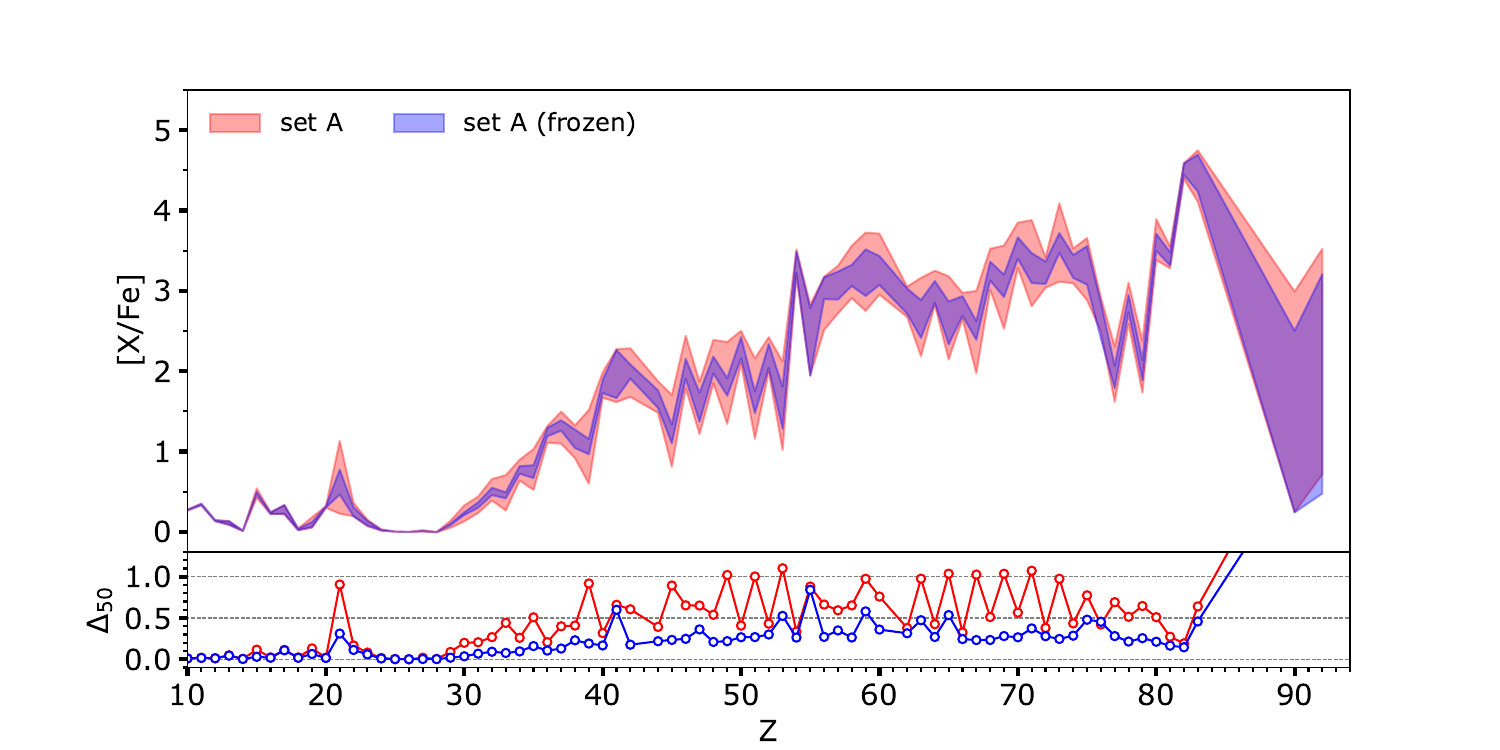}
\caption{Comparison between the surface abundances [X/Fe] and their uncertainties obtained with set A and the ``frozen'' set A where all the \nga rates from Table \ref{Table:main} are fixed at their geometric mean value.}
    \label{fig:frozen}
\end{figure*}

\subsubsection{$^{153}$Sm($n,\gamma$)$^{154}$Sm and $^{217}$Bi($n,\gamma$)$^{218}$Bi}

We explored two other interesting cases that have yet to be experimentally constrained: $^{153}$Sm\nga and $^{217}$Bi\nga. $^{153}$Sm\nga is found to have an impact on the $^{153}$Eu surface abundances {(see Table~\ref{Table:tracer})}, an important observable tracer. The total uncertainty on the $^{153}$Eu i-process production by AGB star amounts to 1.15 dex for set A. We run two additional stellar simulations similarly to the $^{139}$Ba method to quantify the impact of $^{153}$Sm\nga. The resulting uncertainty on $^{153}$Eu stemming directly from the uncertainty of this rate is reduced to 0.69 dex out of the 1.15 dex of the total uncertainty on $^{153}$Eu.

The $^{217}$Bi\nga is also an interesting reaction, as it seems to impact the production of actinides and more particularly $^{232}$Th, $^{235}$U and $^{238}$U.  
In fact, a low $^{217}$Bi\nga rate will favour the $\beta$-decay to a region where $\alpha$-decay strongly dominates. This leads to a cycle denying the production of higher $Z$ nuclei and producing elements in the Pb region \citep[cf. Sect.~3 in][for more details]{Choplin2022letter}. In contrast, a $^{217}$Bi\nga rate larger than the $\beta$-decay rate favours the flux to capture more neutrons, escaping the $\alpha$-decay dominated region producing higher $Z$ actinides, and especially boosting the production of these few long-lived isotopes in this region of the chart of nuclides. 
Using the same method as above, we find that the resulting uncertainty coming from the direct uncertainty of this rate is in fact of 0.99 dex for $^{232}$Th, 1.01 dex for $^{235}$U and 1.01 dex for $^{238}$U. Comparing these values with those given in Table~\ref{Table:main}, clearly $^{217}$Bi\nga plays a key role in the possible enrichment of Th and U by the i-process, but other combinations of uncertain rates also contribute to the $\sim 2.5 - 3$ dex found in the overall surface abundance uncertainty with set A.

\subsubsection{Remaining uncertainties:  combinations of multiples reactions}

We have constrained important reactions in Sect. \ref{Sect:important_reaction} and we investigate in this section other reactions or combinations of reactions that can have an important effect on surface abundances uncertainties. For this purpose, we run another 50 stellar evolution simulations for set A, with the specificity that we froze the reaction rates of all the \nga reactions from Table \ref{Table:main} (roughly 125 out of 868) to their geometric mean value. All the other reactions are once again randomly distributed among their minimum and maximum rates. By doing so, we expect to see if the reactions listed in Table~\ref{Table:main} can account for most of the uncertainties or if the combinations of all the other remaining reactions still have a significant impact.
Figure \ref{fig:frozen} shows the comparison between the set A and the "frozen" set A described above. {With this latter set, the uncertainties are clearly reduced}, especially the odd-$Z$ elements. Most of the uncertainties are now below 0.5 dex, meaning an average 0.5 dex reduction of uncertainties for the odd-$Z$ elements. Large uncertainties still remain for the production of Th and U. This truly underlines the sensitivity of their nucleosynthesis to combinations of multiples reactions and not only to a specific one. For the rest of the nuclei above $Z=40$, all the other reactions contribute on average to a 0.5 dex uncertainty. 
Finally, we note that the range of possible [Ba/Eu] ratios for the frozen set A ranges between $0.13 < \mathrm{[Ba/Eu]} < 0.68$, {\it i.e.} an uncertainty of 0.55 dex, {against 1.35 dex in the standard set A}  (cf. Sect.~\ref{sec:xfe-syst}). 
For the [Ba/La] ratio, we obtain $-0.15 < \mathrm{[Ba/La]} < 0.10$, {\it i.e.} an uncertainty of 0.25 dex, {against 0.98 dex in the standard set A.}

\subsection{Limitations of the method and potential improvements}

The method described in Sect. \ref{Sect:important_reaction} to determine important reactions has of course limitations. The two main problems are to quantify the  impact of each individual reaction and the completeness of the set of reactions we derive. The method that we use here is double-edged: by exploring the abundances variations with random combinations of maximum and minimum rates, we significantly reduce the number of stellar simulations. Indeed, methods often used in other studies \citep[such as][]{denissenkov18,mckay20,denissenkov21} vary randomly individual rates, and hence requires thousands of simulations to extract sensitivity results on all possible cases. This is incredibly CPU-consuming for 1D multi-zone stellar models and hence often imply the use of simpler one-zone models which are not able to capture the complexity of stellar models. By finding first the maximum and minimum rates coming from the statistical uncertainties through the BFMC method, we can in fact perform a complete sensitivity study of a very large number of rates, with a very reduced number of stellar simulations (here $N=50$). The downside of this method, however, is that by varying randomly rates all together, we cannot estimate the direct impact of a specific rate uncertainty on the final abundances. As discussed in Sect. \ref{Sect:direct_impact_reactions}, a solution is to compute stellar models with individual rate variations for each reaction to quantify the direct impact of one reaction on the surface abundances. This means 868 reactions times two rates (maximum/minimum) times the number of nuclear models (two here), meaning a tremendous number of stellar simulations. The second downside of our method is the completeness of the set of important reactions derived. Indeed, our method can relatively easily identify the reactions that have a strong impact on one or multiple isotopic abundances (see Sect. \ref{Sect:important_reaction}). However, we can only extract reactions from isotopic abundances if those are affected mainly by one given reaction, but, if many reactions contribute, we cannot disentangle the various contributions. This was illustrated in Sect. \ref{Sect:direct_impact_reactions} when quantifying the sensitivity of the  $^{139}$Ba\nga reaction. While the direct impact of $^{139}$Ba\nga on the $^{139}$La abundance is of 0.4 dex, the total uncertainty when varying all the 868 unknown \nga rates is of 0.77 dex, and after reducing the uncertainties of the $^{139}$Ba\nga rate, there is still an uncertainty of roughly 0.4 dex. We found the same result for $^{153}$Sm\nga and $^{217}$Bi($n,\gamma$). This shows that the contribution from multiple  reactions impacts significantly the total uncertainty of an isotopic abundance. A perspective to unveil all the multiple reactions that could impact a specific nuclei would be the use of newly developed statistical tools, such as the Global Sensitivity Analysis \citep[see][]{Chatterjee2021,Benesse2022,BenesseThesis}. However, to perform such a statistical analysis, a much larger sample of stellar simulations would be needed. Now that we have showed the stability of the method developed here, we plan to perform such a sensitivity study in a forthcoming work.

\section{Conclusions}


We investigated both the model (systematic) and  parameter (statistical) uncertainties associated with theoretical neutron-capture rates relevant for the i-process nucleosynthesis.  We computed 9 sets of TALYS rates to estimate the correlated model uncertainties and for two of those sets (one based on rather microscopic ingredients, set A, and one that considers more phenomenological models, set B), we applied the BFMC approach to anchor the parameter uncertainties with available experimental data, {\it i.e.} the known neutron-capture rates. Doing so, we obtain for both sets A and B, an estimate of the upper and lower limits to the 868 unknown \nga rates involved in our i-process network. Correlated model and non-correlated parameter uncertainties are globally of the same order of magnitude but can differ quite significantly for the neutron-rich nuclei involved in the i-process.

We determine the impact of these nuclear uncertainties on the surface abundances of a 1~$M_\odot$ [Fe/H]~$= -2.5$ multi-zone stellar model during its early AGB phase which is subject to a proton ingestion event followed by an i-process nucleosynthesis with neutron densities of $\approx 10^{15}$~cm$^{-3}$. 

By considering 9 different nuclear models with different combinations of NLD and PSF models, we found that the correlated model uncertainties lead to surface abundance uncertainties for $Z \geq 40$ elements to range between 0.5 and 1.0 dex, with the exception of Th and U for which the uncertainty rises to about 3 dex. Due to the correlated nature of these uncertainties, abundance ratios remain much better constrained.

Similarly, the non-correlated parameter uncertainties give rise to AGB surface abundance uncertainties for $Z \geq 40$ elements also to range between 0.5 and 1.0 dex, though odd-$Z$ elements display significantly higher uncertainties due to their reduced number of stable isotopes and higher sensitivity to specific rates. Interestingly, the impact of uncorrelated parameter uncertainties obtained with two different nuclear models is found to be rather similar. However, the uncorrelated nature of these nuclear uncertainties has a significant impact on abundance ratios.

Both source of nuclear uncertainties have an important impact on the predicted abundance of potential observable tracers such as Eu and La.  Interestingly, the choice of the BFMC $\chi^2$ estimator mainly impacts the abundances of odd-$Z$ elements. 
We reconfirm that the production of actinides is possible in this i-process site \citep{Choplin2022letter}, but remains highly sensitive to both the parameter and model uncertainties. The resulting abundance uncertainties of actinides are in fact much larger than for any other species and underline the need for improved nuclear predictions for the neutron-capture rates of $Z \geq 82$ nuclei along the i-process path.  

We developed a method to estimate surface abundance uncertainties in 1D multi-zone models and extract the impacting reactions affecting one or multiple isotopic abundances. We found roughly 125 \nga reactions, including 25 with high impact on elemental surface abundances uncertainties.
Interestingly, more than 30 \nga reactions have medium to high impact on the surface abundance of elements that are usually taken as observable tracers of the i-process nucleosynthesis in CEMP stars. One of these reactions, $^{139}$Ba($n,\gamma$)$^{140}$Ba, has  been recently experimentally constrained (M\"ucher \& Spyrou, priv. comm). 
We found that a 85\% reduction of the uncertainties on this rate leads to a reduction of 0.4 dex out of the total 0.77 dex uncertainty on the La production.

Finally we showed that even by greatly reducing the uncertainties of the $\sim 125$ main \nga reactions impacting the AGB surface abundances, there are still uncertainties coming from complex combinations of multiple rate uncertainties. This means that the reductions of rate uncertainties, even for the reactions not listed here, might have a global impact on the surface abundance uncertainties. Moreover, new measurement of neutron captures on neutron-rich nuclei would also help to improve the theoretical nuclear models, in particular for the description of NLDs and PSFs.
{Our analysis of the i-process uncertainty was conducted on a a specific stellar model. An interesting perspective is to apply our method to PIE in AGB stars of different masses and different metallicity but also to investigate other sites such as RAWD \citep{denissenkov17,denissenkov21}. Further developments to use Global Sensitivity Analysis techniques \citep{Benesse2022} could also help identify complete sets of relevant reactions.}

\begin{acknowledgements}
SM and SG has received support from the European Union (ChECTEC-INFRA, project no. 101008324). 
LS and SG are senior F.R.S.-FNRS research associates.
A.C. is a Postdoctoral Researcher of the Fonds de la Recherche Scientifique – FNRS.
Figure 12 is an adaptation from \url{https://github.com/kmiernik/Chart-of-nuclides-drawer}.
\end{acknowledgements}

\bibliographystyle{aa} 
\bibliography{references.bib} 

\afterpage{
\clearpage
\onecolumn
\input{Table_total_new.tex}
\twocolumn
\clearpage
}

\afterpage{
\clearpage
\onecolumn
\input{Table_tracer_new.tex}
\twocolumn
\clearpage
}
\newpage

\end{document}

%% file: Table_total_new.tex
\begin{longtable}[c]{lrrrccccc}

\caption{List of non-experimental \nga reactions and the impact of their nuclear uncertainty on the surface abundances in our low-mass low-metallicity AGB star model.
{The reported quantities are the nucleus properties (columns 1, 2 and 3), the contributing range of the concerned isotope to the total elemental abundance (column 4), the ratio between the maximum and the minimum abundances (in log: $\Delta_{50}$) for sets A and B (columns 5 and 6), the reaction directly impacting its abundance (column 7) and the ratio between the maximum and minimum rates of the concerned \nga reaction (columns 8 and 9).}
\label{Table:main}}\\
\toprule
\toprule
Element &  Z &   A & Iso. Frac.& \multicolumn{2}{|c|}{Surf. abund. uncertainty (in log)} &Reaction & \multicolumn{2}{|c}{$\langle\sigma \rangle_{\rm max}~/~\langle\sigma \rangle_{\rm min}$}\\
        &    &     &          & ~~~set A~ & set B & & set A & set B \\
\midrule
\endhead
\midrule
\multicolumn{9}{r}{{Continued on next page}} \\
\midrule
\endfoot

\bottomrule
\endlastfoot
 U & 92 & 235 &               8-51\% &                       2.90 &                       1.37 & $^{217}$Bi(n,$\gamma$) &         57.2 &         10.0 \\
 U & 92 & 238 &              49-92\% &                       2.87 &                       1.80 & $^{217}$Bi(n,$\gamma$) &         57.2 &         10.0 \\
Th & 90 & 232 &             100\% &                       2.75 &                       1.36 & $^{217}$Bi(n,$\gamma$) &         57.2 &         10.0 \\
Dy & 66 & 160 &               0-1\% &                       1.66 &                       0.91 & $^{160}$Tb(n,$\gamma$) &          7.5 &          3.2 \\
Dy & 66 & 160 &               0-1\% &                       1.66 &                       0.91 & $^{159}$Gd(n,$\gamma$) &         12.0 &          6.5 \\
Gd & 64 & 154 &               $<$0.5\% &                       1.39 &                       1.02 & $^{153}$Sm(n,$\gamma$) &         12.5 &          5.5 \\
Ba & 56 & 137 &               4-85\% &                       1.34 &                       1.95 & $^{137}$Xe(n,$\gamma$) &         11.6 &          8.4 \\
Ba & 56 & 137 &               4-85\% &                       1.34 &                       1.95 & $^{137}$Cs(n,$\gamma$) &         15.4 &         78.4 \\
Pb & 82 & 204 &               $<$0.5\% &                       1.31 &                       1.55 & $^{203}$Hg(n,$\gamma$) &          6.3 &          9.8 \\
Sm & 62 & 150 &               0-4\% &                       1.31 &                       1.08 & $^{149}$Nd(n,$\gamma$) &          7.3 &          5.4 \\
Xe & 54 & 130 &               $<$0.5\% &                       1.30 &                       0.78 &  $^{130}$I(n,$\gamma$) &          8.6 &          2.7 \\
Sb & 51 & 121 &              10-72\% &                       1.24 &                       0.95 & $^{121}$Sn(n,$\gamma$) &          9.4 &          4.5 \\
Sm & 62 & 148 &               $<$0.5\% &                       1.20 &                       1.01 & $^{148}$Pm(n,$\gamma$) &          8.9 &          3.1 \\
Sm & 62 & 148 &               $<$0.5\% &                       1.20 &                       1.01 & $^{147}$Nd(n,$\gamma$) &         10.5 &          9.5 \\
Mo & 42 &  95 &               2-66\% &                       1.18 &                       1.19 &  $^{95}$Zr(n,$\gamma$) &         11.5 &         11.8 \\
Os & 76 & 188 &              30-68\% &                       1.17 &                       0.62 &  $^{188}$W(n,$\gamma$) &          8.6 &          3.2 \\
Ba & 56 & 136 &               0-6\% &                       1.17 &                       1.58 & $^{135}$Xe(n,$\gamma$) &          5.1 &          7.6 \\
Ba & 56 & 136 &               0-6\% &                       1.17 &                       1.58 & $^{136}$Cs(n,$\gamma$) &          6.8 &         14.2 \\
Eu & 63 & 153 &              15-82\% &                       1.15 &                       0.88 & $^{153}$Sm(n,$\gamma$) &         12.5 &          5.5 \\
Sm & 62 & 147 &               6-51\% &                       1.14 &                       1.17 & $^{147}$Pr(n,$\gamma$) &         11.9 &          9.8 \\
Sm & 62 & 147 &               6-51\% &                       1.14 &                       1.17 & $^{147}$Nd(n,$\gamma$) &         10.5 &          9.5 \\
Cd & 48 & 111 &               2-12\% &                       1.10 &                       0.71 & $^{111}$Pd(n,$\gamma$) &          6.8 &          5.1 \\
 I & 53 & 127 &             100\% &                       1.10 &                       1.18 & $^{127}$Sb(n,$\gamma$) &         10.7 &         12.4 \\
Nd & 60 & 143 &               2-36\% &                       1.10 &                       0.99 & $^{143}$Ce(n,$\gamma$) &         12.6 &          8.9 \\
Nd & 60 & 144 &              27-79\% &                       1.07 &                       0.85 & $^{144}$Ce(n,$\gamma$) &          7.2 &          4.7 \\
Lu & 71 & 175 &             100\% &                       1.07 &                       0.81 & $^{175}$Yb(n,$\gamma$) &          8.7 &          5.6 \\
Hg & 80 & 200 &               9-46\% &                       1.07 &                       0.84 & $^{200}$Pt(n,$\gamma$) &         11.3 &          5.6 \\
Sm & 62 & 149 &               4-30\% &                       1.07 &                       0.84 & $^{149}$Nd(n,$\gamma$) &          7.3 &          5.4 \\
Pd & 46 & 106 &              27-75\% &                       1.07 &                       0.93 & $^{106}$Ru(n,$\gamma$) &          8.0 &          4.3 \\
Nd & 60 & 145 &               1-18\% &                       1.05 &                       0.91 & $^{145}$Pr(n,$\gamma$) &         12.0 &          8.7 \\
Pt & 78 & 194 &              26-63\% &                       1.04 &                       0.68 & $^{194}$Os(n,$\gamma$) &          8.7 &          3.4 \\
Tb & 65 & 159 &             100\% &                       1.04 &                       0.86 & $^{159}$Gd(n,$\gamma$) &         12.0 &          6.5 \\
Tm & 69 & 169 &             100\% &                       1.04 &                       0.89 & $^{169}$Er(n,$\gamma$) &          9.6 &          5.3 \\
Eu & 63 & 151 &              18-85\% &                       1.04 &                       0.83 & $^{151}$Pm(n,$\gamma$) &          8.6 &          6.1 \\
Ho & 67 & 165 &             100\% &                       1.03 &                       0.78 & $^{165}$Dy(n,$\gamma$) &         11.1 &          6.3 \\
In & 49 & 115 &             100\% &                       1.02 &                       0.82 & $^{115}$Cd(n,$\gamma$) &          7.6 &          4.4 \\
Sn & 50 & 115 &               $<$0.5\% &                       1.02 &                       0.82 & $^{115}$Cd(n,$\gamma$) &          7.6 &          4.4 \\
Xe & 54 & 131 &               0-2\% &                       1.01 &                       1.06 &  $^{131}$I(n,$\gamma$) &          8.1 &          9.2 \\
Yb & 70 & 172 &              16-58\% &                       1.01 &                       0.76 & $^{172}$Er(n,$\gamma$) &          9.3 &          3.7 \\
Lu & 71 & 176 &               $<$0.5\% &                       1.01 &                       0.86 & $^{175}$Yb(n,$\gamma$) &          8.7 &          5.6 \\
Yb & 70 & 171 &               3-30\% &                       1.01 &                       0.89 & $^{171}$Er(n,$\gamma$) &          9.1 &          6.5 \\
Hf & 72 & 176 &               $<$0.5\% &                       1.00 &                       0.86 & $^{175}$Yb(n,$\gamma$) &          8.7 &          5.6 \\
Er & 68 & 166 &              23-63\% &                       1.00 &                       0.75 & $^{166}$Dy(n,$\gamma$) &          7.1 &          3.5 \\
Hg & 80 & 199 &               2-16\% &                       0.99 &                       0.76 & $^{199}$Au(n,$\gamma$) &          5.6 &          8.0 \\
Hf & 72 & 177 &               5-29\% &                       0.99 &                       0.84 & $^{177}$Yb(n,$\gamma$) &          9.3 &          5.9 \\
Yb & 70 & 173 &               3-21\% &                       0.99 &                       0.78 & $^{173}$Tm(n,$\gamma$) &          9.9 &          7.2 \\
 W & 74 & 184 &              15-61\% &                       0.99 &                       0.64 & $^{184}$Hf(n,$\gamma$) &         10.2 &          3.6 \\
Ta & 73 & 181 &             100\% &                       0.98 &                       1.00 & $^{181}$Hf(n,$\gamma$) &          7.5 &          6.4 \\
Gd & 64 & 156 &              11-58\% &                       0.97 &                       0.85 & $^{156}$Sm(n,$\gamma$) &          7.5 &          3.7 \\
Pr & 59 & 141 &             100\% &                       0.97 &                       0.96 & $^{141}$La(n,$\gamma$) &         11.1 &         10.0 \\
Tl & 81 & 203 &               9-58\% &                       0.96 &                       1.05 & $^{203}$Hg(n,$\gamma$) &          6.3 &          9.8 \\
Ag & 47 & 109 &              30-78\% &                       0.96 &                       0.86 & $^{109}$Pd(n,$\gamma$) &          6.2 &          4.4 \\
Ir & 77 & 193 &              39-90\% &                       0.93 &                       0.77 & $^{193}$Os(n,$\gamma$) &          6.5 &          4.1 \\
 Y & 39 &  89 &             100\% &                       0.92 &                       0.78 &  $^{89}$Sr(n,$\gamma$) &          7.6 &          8.4 \\
Re & 75 & 185 &              14-78\% &                       0.91 &                       0.78 & $^{185}$Ta(n,$\gamma$) &         10.5 &          6.0 \\
Sc & 21 &  45 &             100\% &                       0.91 &                       0.97 &  $^{45}$Ca(n,$\gamma$) &         10.1 &          9.7 \\
Sn & 50 & 117 &               1-4\% &                       0.90 &                       0.72 & $^{117}$Cd(n,$\gamma$) &          6.4 &          4.9 \\
Rh & 45 & 103 &             100\% &                       0.89 &                       1.07 & $^{103}$Ru(n,$\gamma$) &          6.8 &          5.8 \\
Gd & 64 & 157 &               4-23\% &                       0.88 &                       0.89 & $^{157}$Eu(n,$\gamma$) &          9.3 &          7.4 \\
Cs & 55 & 133 &             100\% &                       0.88 &                       1.06 &  $^{133}$I(n,$\gamma$) &          6.5 &          9.5 \\
Zr & 40 &  91 &               2-28\% &                       0.88 &                       1.12 &  $^{91}$Sr(n,$\gamma$) &          8.2 &         11.9 \\
Dy & 66 & 161 &               5-26\% &                       0.87 &                       0.76 & $^{161}$Tb(n,$\gamma$) &          9.4 &          6.0 \\
Sb & 51 & 123 &              28-90\% &                       0.86 &                       1.01 & $^{123}$Sn(n,$\gamma$) &          4.0 &          5.4 \\
Er & 68 & 167 &               3-23\% &                       0.86 &                       0.70 & $^{167}$Ho(n,$\gamma$) &          9.6 &          7.2 \\
Pd & 46 & 105 &               2-19\% &                       0.86 &                       0.73 & $^{105}$Ru(n,$\gamma$) &          8.4 &          7.0 \\
Re & 75 & 187 &              22-86\% &                       0.83 &                       0.73 &  $^{187}$W(n,$\gamma$) &          4.4 &          4.6 \\
Mo & 42 &  97 &               8-72\% &                       0.83 &                       1.39 &  $^{97}$Zr(n,$\gamma$) &          5.4 &         21.0 \\
 W & 74 & 183 &               7-45\% &                       0.82 &                       0.75 & $^{183}$Hf(n,$\gamma$) &          7.8 &          7.8 \\
Gd & 64 & 155 &               3-14\% &                       0.81 &                       0.63 & $^{155}$Sm(n,$\gamma$) &         13.6 &          8.0 \\
Pt & 78 & 195 &               3-15\% &                       0.81 &                       0.60 & $^{195}$Ir(n,$\gamma$) &          7.3 &          5.0 \\
Hg & 80 & 202 &              17-61\% &                       0.79 &                       0.69 & $^{202}$Pt(n,$\gamma$) &          6.0 &          8.6 \\
Xe & 54 & 129 &               0-1\% &                       0.79 &                       0.89 & $^{129}$Sb(n,$\gamma$) &          6.4 &         12.4 \\
Ce & 58 & 140 &              27-90\% &                       0.78 &                       0.86 & $^{140}$Ba(n,$\gamma$) &          3.2 &          4.3 \\
Br & 35 &  79 &              31-75\% &                       0.77 &                       0.62 &  $^{79}$Se(n,$\gamma$) &          7.3 &          5.7 \\
Zn & 30 &  70 &               3-20\% &                       0.74 &                       0.87 &  $^{70}$Zn(n,$\gamma$) &          2.8 &          3.0 \\
Cd & 48 & 112 &              23-55\% &                       0.74 &                       0.75 & $^{112}$Pd(n,$\gamma$) &          4.5 &          3.6 \\
Ru & 44 & 101 &               5-14\% &                       0.71 &                       0.66 & $^{101}$Mo(n,$\gamma$) &          9.4 &         10.3 \\
Ti & 22 &  47 &               8-38\% &                       0.69 &                       0.86 &  $^{47}$Ca(n,$\gamma$) &          7.5 &         10.0 \\
Nd & 60 & 142 &               $<$0.5\% &                       0.68 &                       0.30 & $^{142}$Pr(n,$\gamma$) &         12.6 &          3.1 \\
Gd & 64 & 158 &               9-25\% &                       0.68 &                       0.43 & $^{158}$Eu(n,$\gamma$) &          8.9 &          3.6 \\
Cd & 48 & 110 &               0-1\% &                       0.67 &                       0.58 & $^{109}$Pd(n,$\gamma$) &          6.2 &          4.4 \\
Pb & 82 & 206 &               4-28\% &                       0.66 &                       0.68 & $^{210}$Pb(n,$\gamma$) &          5.9 &         15.1 \\
Nb & 41 &  93 &             100\% &                       0.66 &                       0.75 &   $^{93}$Y(n,$\gamma$) &          6.3 &          7.1 \\
Xe & 54 & 132 &               1-5\% &                       0.66 &                       1.02 & $^{132}$Te(n,$\gamma$) &          2.2 &          6.1 \\
Ba & 56 & 138 &               9-80\% &                       0.66 &                       0.52 & $^{138}$Cs(n,$\gamma$) &          7.0 &          4.8 \\
Os & 76 & 189 &               4-22\% &                       0.65 &                       0.59 & $^{189}$Re(n,$\gamma$) &          3.8 &          4.6 \\
Au & 79 & 197 &             100\% &                       0.65 &                       0.77 & $^{197}$Pt(n,$\gamma$) &          3.4 &          4.5 \\
Bi & 83 & 209 &             100\% &                       0.64 &                       0.73 & $^{209}$Pb(n,$\gamma$) &          7.1 &         10.2 \\
Hf & 72 & 179 &               6-22\% &                       0.64 &                       0.57 & $^{179}$Lu(n,$\gamma$) &          9.6 &          6.6 \\
Se & 34 &  77 &               4-22\% &                       0.63 &                       0.75 &  $^{77}$Ge(n,$\gamma$) &          7.2 &          8.2 \\
Te & 52 & 124 &               $<$0.5\% &                       0.62 &                       0.77 & $^{123}$Sn(n,$\gamma$) &          4.0 &          5.4 \\
Zr & 40 &  90 &              10-43\% &                       0.62 &                       0.73 &  $^{90}$Sr(n,$\gamma$) &          2.8 &          3.0 \\
Zr & 40 &  92 &               7-25\% &                       0.61 &                       0.57 &  $^{92}$Sr(n,$\gamma$) &          4.0 &          3.9 \\
Kr & 36 &  82 &               1-3\% &                       0.61 &                       0.30 &  $^{82}$Br(n,$\gamma$) &          7.6 &          2.3 \\
Ir & 77 & 191 &              10-61\% &                       0.60 &                       0.65 & $^{191}$Os(n,$\gamma$) &          2.9 &          3.2 \\
Te & 52 & 122 &               $<$0.5\% &                       0.60 &                       0.29 & $^{121}$Sn(n,$\gamma$) &          9.4 &          4.5 \\
La & 57 & 139 &             100\% &                       0.59 &                       0.77 & $^{139}$Ba(n,$\gamma$) &         10.3 &          9.0 \\
Ge & 32 &  72 &              18-49\% &                       0.59 &                       0.54 &  $^{72}$Zn(n,$\gamma$) &          3.1 &          2.7 \\
Hf & 72 & 178 &              20-44\% &                       0.58 &                       0.48 & $^{178}$Yb(n,$\gamma$) &          8.1 &          3.8 \\
Te & 52 & 130 &              17-41\% &                       0.58 &                       0.67 & $^{130}$Sb(n,$\gamma$) &          7.1 &         10.9 \\
Pt & 78 & 192 &               $<$0.5\% &                       0.56 &                       0.61 & $^{191}$Os(n,$\gamma$) &          2.9 &          3.2 \\
Sn & 50 & 118 &               9-15\% &                       0.55 &                       0.50 & $^{118}$Cd(n,$\gamma$) &          5.5 &          3.5 \\
Ce & 58 & 142 &              10-73\% &                       0.54 &                       0.51 & $^{142}$La(n,$\gamma$) &          7.8 &          3.4 \\
Os & 76 & 190 &              10-33\% &                       0.54 &                       0.39 &  $^{190}$W(n,$\gamma$) &         11.1 &          3.7 \\
Ca & 20 &  43 &               0-1\% &                       0.54 &                       0.57 &   $^{43}$K(n,$\gamma$) &          3.7 &          4.4 \\
Se & 34 &  82 &              10-54\% &                       0.54 &                       0.95 &  $^{82}$Se(n,$\gamma$) &          1.7 &          3.6 \\
Sn & 50 & 116 &               $<$0.5\% &                       0.53 &                       0.34 & $^{115}$Cd(n,$\gamma$) &          7.6 &          4.4 \\
Dy & 66 & 162 &              16-34\% &                       0.52 &                       0.40 & $^{162}$Tb(n,$\gamma$) &         10.4 &          4.0 \\
Pb & 82 & 207 &               3-17\% &                       0.52 &                       0.51 & $^{211}$Bi(n,$\gamma$) &          8.1 &         13.5 \\
Hg & 80 & 201 &               3-11\% &                       0.52 &                       0.49 & $^{201}$Au(n,$\gamma$) &         10.9 &         10.6 \\
Sn & 50 & 119 &               2-4\% &                       0.51 &                       0.46 & $^{118}$Cd(n,$\gamma$) &          5.5 &          3.5 \\
Rb & 37 &  87 &              57-87\% &                       0.50 &                       0.65 &  $^{87}$Kr(n,$\gamma$) &          6.7 &         10.8 \\
 W & 74 & 186 &              22-51\% &                       0.50 &                       0.31 & $^{186}$Ta(n,$\gamma$) &          7.8 &          3.1 \\
Yb & 70 & 174 &               9-27\% &                       0.50 &                       0.39 & $^{174}$Tm(n,$\gamma$) &          8.8 &          3.6 \\
Sr & 38 &  86 &               1-4\% &                       0.49 &                       0.31 &  $^{86}$Rb(n,$\gamma$) &          5.3 &          2.2 \\
Ge & 32 &  73 &               5-18\% &                       0.49 &                       0.55 &  $^{73}$Ga(n,$\gamma$) &          4.1 &          5.4 \\
Ba & 56 & 134 &               $<$0.5\% &                       0.49 &                       0.55 & $^{134}$Cs(n,$\gamma$) &          8.0 &          6.3 \\
Zr & 40 &  94 &               8-24\% &                       0.49 &                       0.48 &   $^{94}$Y(n,$\gamma$) &         10.0 &          3.7 \\
Zn & 30 &  67 &               4-9\% &                       0.47 &                       0.43 &  $^{67}$Cu(n,$\gamma$) &          3.9 &          5.4 \\
Se & 34 &  78 &              11-37\% &                       0.46 &                       0.34 &  $^{78}$Se(n,$\gamma$) &          2.9 &          2.9 \\
Pd & 46 & 104 &               0-1\% &                       0.46 &                       0.70 & $^{103}$Ru(n,$\gamma$) &          6.8 &          5.8 \\
Mo & 42 &  98 &              11-35\% &                       0.46 &                       0.56 &  $^{97}$Zr(n,$\gamma$) &          5.4 &         21.0 \\
Er & 68 & 168 &               9-32\% &                       0.45 &                       0.32 & $^{168}$Ho(n,$\gamma$) &          9.8 &          3.8 \\
Pt & 78 & 196 &              12-29\% &                       0.45 &                       0.30 & $^{196}$Os(n,$\gamma$) &         12.8 &          4.8 \\
As & 33 &  75 &             100\% &                       0.44 &                       0.42 &  $^{75}$Ge(n,$\gamma$) &          6.9 &          8.7 \\
Hf & 72 & 180 &              28-57\% &                       0.44 &                       0.36 & $^{180}$Lu(n,$\gamma$) &         10.3 &          3.7 \\
Ru & 44 &  99 &              14-27\% &                       0.43 &                       0.53 &  $^{97}$Zr(n,$\gamma$) &          5.4 &         21.0 \\
Sr & 38 &  88 &              95-99\% &                       0.42 &                       0.45 &  $^{88}$Kr(n,$\gamma$) &          3.4 &          3.5 \\
Se & 34 &  76 &               2-6\% &                       0.42 &                       0.30 &  $^{75}$Ge(n,$\gamma$) &          6.9 &          8.7 \\
Dy & 66 & 163 &               9-21\% &                       0.41 &                       0.45 & $^{163}$Tb(n,$\gamma$) &         10.0 &          8.5 \\
Mo & 42 & 100 &               9-32\% &                       0.41 &                       0.51 &  $^{97}$Zr(n,$\gamma$) &          5.4 &         21.0 \\
Kr & 36 &  83 &               3-8\% &                       0.39 &                       0.49 &  $^{83}$Br(n,$\gamma$) &          8.1 &          5.2 \\
 K & 19 &  41 &               8-26\% &                       0.36 &                       0.50 &  $^{41}$Ar(n,$\gamma$) &          5.8 &          9.6 \\
Nd & 60 & 150 &               3-13\% &                       0.34 &                       0.38 & $^{149}$Nd(n,$\gamma$) &          7.3 &          5.4 \\
Xe & 54 & 128 &               $<$0.5\% &                       0.32 &                       0.50 & $^{127}$Sb(n,$\gamma$) &         10.7 &         12.4 \\
Os & 76 & 192 &              13-29\% &                       0.31 &                       0.29 &  $^{190}$W(n,$\gamma$) &         11.1 &          3.7 \\
Zn & 30 &  66 &              28-48\% &                       0.29 &                       0.30 &  $^{66}$Ni(n,$\gamma$) &          2.4 &          2.4 \\
 S & 16 &  33 &               2-9\% &                       0.27 &                       0.64 &   $^{33}$P(n,$\gamma$) &          1.8 &          3.1 \\
Ga & 31 &  69 &              46-66\% &                       0.26 &                       0.28 &  $^{69}$Zn(n,$\gamma$) &          4.9 &          7.0 \\
Br & 35 &  81 &              25-69\% &                       0.21 &                       0.35 &  $^{81}$Se(n,$\gamma$) &          3.8 &          8.8 \\
Cu & 29 &  65 &              33-47\% &                       0.20 &                       0.24 &  $^{65}$Ni(n,$\gamma$) &          5.4 &          9.0 \\
Ge & 32 &  70 &               6-14\% &                       0.19 &                       0.20 &  $^{69}$Zn(n,$\gamma$) &          4.9 &          7.0 \\
Mo & 42 &  96 &               0-1\% &                       0.17 &                       0.12 &  $^{95}$Zr(n,$\gamma$) &         11.5 &         11.8 \\
Mo & 42 &  96 &               0-1\% &                       0.17 &                       0.12 &  $^{96}$Nb(n,$\gamma$) &         11.9 &          3.3 \\
Cl & 17 &  35 &              51-80\% &                       0.14 &                       0.43 &   $^{35}$S(n,$\gamma$) &          2.5 &          8.5 \\
 K & 19 &  39 &              74-92\% &                       0.12 &                       0.16 &  $^{39}$Ar(n,$\gamma$) &         10.4 &         20.4 \\
 P & 15 &  31 &             100\% &                       0.12 &                       0.17 &  $^{31}$Si(n,$\gamma$) &          2.3 &          4.8 \\
Er & 68 & 164 &               $<$0.5\% &                       0.09 &                       0.05 & $^{164}$Ho(n,$\gamma$) &          6.9 &          2.7 \\
Ar & 18 &  38 &              17-21\% &                       0.09 &                       0.10 &  $^{38}$Ar(n,$\gamma$) &          2.4 &          2.7 \\
 S & 16 &  34 &              21-29\% &                       0.03 &                       0.13 &  $^{31}$Si(n,$\gamma$) &          2.3 &          4.8 \\
 S & 16 &  34 &              21-29\% &                       0.03 &                       0.13 &  $^{32}$Si(n,$\gamma$) &          1.3 &          4.2 \\
Ca & 20 &  42 &              47-49\% &                       0.02 &                       0.02 &  $^{41}$Ar(n,$\gamma$) &          5.8 &          9.6 \\
 S & 16 &  32 &              56-68\% &                       0.02 &                       0.08 &  $^{32}$Si(n,$\gamma$) &          1.3 &          4.2 \\
 S & 16 &  32 &              56-68\% &                       0.02 &                       0.08 &  $^{31}$Si(n,$\gamma$) &          2.3 &          4.8 \\
Ti & 22 &  48 &              31-59\% &                       0.02 &                       0.02 &  $^{47}$Ca(n,$\gamma$) &          7.5 &         10.0 \\
Te & 52 & 123 &               $<$0.5\% &                       0.02 &                       0.01 & $^{121}$Sn(n,$\gamma$) &          9.4 &          4.5 \\
Co & 27 &  59 &             100\% &                       0.02 &                       0.04 &  $^{59}$Fe(n,$\gamma$) &          3.9 &         11.9 \\
\bottomrule

\end{longtable}

%% file: Table_tracer_new.tex
\begin{longtable}[c]{lrrrccccc}
\caption{Same as Table \ref{Table:main} for isotopes of the main observable heavy elements in CEMP stars, {\it i.e.} i-process nucleosynthesis tracers. 
\label{Table:tracer} 
}\\
\toprule
\toprule
Element &  Z &   A & Iso. Frac.& \multicolumn{2}{|c|}{Surface abund. uncertainty (in log)}& Reaction & \multicolumn{2}{|c}{$\langle\sigma \rangle_{\rm max}~/~\langle\sigma \rangle_{\rm min}$}\\
        &    &     &          &  ~~set A & set B &  & set A & set B \\
\midrule
\endhead
\midrule
\multicolumn{9}{r}{{Continued on next page}} \\
\midrule
\endfoot
\bottomrule
\endlastfoot
Sr & 38 &  86 &               1-4\% &                       0.49 &                       0.31 &  $^{86}$Rb(n,$\gamma$) &        5.3 &          2.2 \\
Sr & 38 &  87 &               0-1\% &                       0.01 &                       0.00 &  $^{86}$Rb(n,$\gamma$) &        5.3 &          2.2 \\
Sr & 38 &  88 &              95-99\% &                       0.42 &                       0.45 &  $^{88}$Kr(n,$\gamma$) &       3.4 &          3.5 \\
\midrule
 Y & 39 &  89 &             100\% &                       0.92 &                       0.78 &  $^{89}$Sr(n,$\gamma$) &          7.6 &          8.4 \\
\midrule
Zr & 40 &  90 &              10-43\% &                       0.62 &                       0.73 &  $^{90}$Sr(n,$\gamma$) &       2.8 &          3.0 \\
Zr & 40 &  91 &               2-28\% &                       0.88 &                       1.12 &  $^{91}$Sr(n,$\gamma$) &       8.2 &         11.9 \\
Zr & 40 &  92 &               7-25\% &                       0.61 &                       0.57 &  $^{92}$Sr(n,$\gamma$) &       4.0 &          3.9 \\
Zr & 40 &  94 &               8-24\% &                       0.49 &                       0.48 &   $^{94}$Y(n,$\gamma$) &      10.0 &          3.7 \\
\midrule
Ba & 56 & 134 &               $<$0.5\% &                       0.49 &                       0.55 & $^{134}$Cs(n,$\gamma$) &     8.0 &          6.3 \\
Ba & 56 & 136 &               0-6\% &                       1.17 &                       1.58 & $^{135}$Xe(n,$\gamma$) &        5.1 &          7.6 \\
Ba & 56 & 136 &               0-6\% &                       1.17 &                       1.58 & $^{136}$Cs(n,$\gamma$) &        6.8 &         14.2 \\
Ba & 56 & 137 &               4-85\% &                       1.34 &                       1.95 & $^{137}$Xe(n,$\gamma$) &      11.6 &          8.4 \\
Ba & 56 & 137 &               4-85\% &                       1.34 &                       1.95 & $^{137}$Cs(n,$\gamma$) &      15.4 &         78.4 \\
Ba & 56 & 138 &               9-80\% &                       0.66 &                       0.52 & $^{138}$Cs(n,$\gamma$) &       7.0 &          4.8 \\
\midrule
La & 57 & 139 &             100\% &                       0.59 &                       0.77 & $^{139}$Ba(n,$\gamma$) &         10.3 &          9.0 \\
\midrule
Ce & 58 & 140 &              27-90\% &                       0.78 &                       0.86 & $^{140}$Ba(n,$\gamma$) &       3.2 &          4.3 \\
Ce & 58 & 142 &              10-73\% &                       0.54 &                       0.51 & $^{142}$La(n,$\gamma$) &       7.8 &          3.4 \\
\midrule
Pr & 59 & 141 &             100\% &                       0.97 &                       0.96 & $^{141}$La(n,$\gamma$) &         11.1 &         10.0 \\
\midrule
Nd & 60 & 142 &               $<$0.5\% &                       0.68 &                       0.30 & $^{142}$Pr(n,$\gamma$) &    12.6 &          3.1 \\
Nd & 60 & 143 &               2-36\% &                       1.10 &                       0.99 & $^{143}$Ce(n,$\gamma$) &      12.6 &          8.9 \\
Nd & 60 & 144 &              27-79\% &                       1.07 &                       0.85 & $^{144}$Ce(n,$\gamma$) &       7.2 &          4.7 \\
Nd & 60 & 145 &               1-18\% &                       1.05 &                       0.91 & $^{145}$Pr(n,$\gamma$) &      12.0 &          8.7 \\
Nd & 60 & 150 &               3-13\% &                       0.34 &                       0.38 & $^{149}$Nd(n,$\gamma$) &       7.3 &          5.4 \\
\midrule
Sm & 62 & 147 &               6-51\% &                       1.14 &                       1.17 & $^{147}$Pr(n,$\gamma$) &      11.9 &          9.8 \\
Sm & 62 & 147 &               6-51\% &                       1.14 &                       1.17 & $^{147}$Nd(n,$\gamma$) &      10.5 &          9.5 \\
Sm & 62 & 148 &               $<$0.5\% &                       1.20 &                       1.01 & $^{148}$Pm(n,$\gamma$) &     8.9 &          3.1 \\
Sm & 62 & 148 &               $<$0.5\% &                       1.20 &                       1.01 & $^{147}$Nd(n,$\gamma$) &    10.5 &          9.5 \\
Sm & 62 & 149 &               4-30\% &                       1.07 &                       0.84 & $^{149}$Nd(n,$\gamma$) &       7.3 &          5.4 \\
Sm & 62 & 150 &               0-4\% &                       1.31 &                       1.08 & $^{149}$Nd(n,$\gamma$) &        7.3 &          5.4 \\
\midrule
Eu & 63 & 151 &              18-85\% &                       1.04 &                       0.83 & $^{151}$Pm(n,$\gamma$) &       8.6 &          6.1 \\
Eu & 63 & 153 &              15-82\% &                       1.15 &                       0.88 & $^{153}$Sm(n,$\gamma$) &      12.5 &          5.5 \\
\midrule
Gd & 64 & 154 &               $<$0.5\% &                       1.39 &                       1.02 & $^{153}$Sm(n,$\gamma$) &    12.5 &          5.5 \\
Gd & 64 & 155 &               3-14\% &                       0.81 &                       0.63 & $^{155}$Sm(n,$\gamma$) &      13.6 &          8.0 \\
Gd & 64 & 156 &              11-58\% &                       0.97 &                       0.85 & $^{156}$Sm(n,$\gamma$) &       7.5 &          3.7 \\
Gd & 64 & 157 &               4-23\% &                       0.88 &                       0.89 & $^{157}$Eu(n,$\gamma$) &       9.3 &          7.4 \\
Gd & 64 & 158 &               9-25\% &                       0.68 &                       0.43 & $^{158}$Eu(n,$\gamma$) &       8.9 &          3.6 \\
\midrule
Dy & 66 & 160 &               0-1\% &                       1.66 &                       0.91 & $^{160}$Tb(n,$\gamma$) &        7.5 &          3.2 \\
Dy & 66 & 160 &               0-1\% &                       1.66 &                       0.91 & $^{159}$Gd(n,$\gamma$) &       12.0 &          6.5 \\
Dy & 66 & 161 &               5-26\% &                       0.87 &                       0.76 & $^{161}$Tb(n,$\gamma$) &       9.4 &          6.0 \\
Dy & 66 & 162 &              16-34\% &                       0.52 &                       0.40 & $^{162}$Tb(n,$\gamma$) &      10.4 &          4.0 \\
Dy & 66 & 163 &               9-21\% &                       0.41 &                       0.45 & $^{163}$Tb(n,$\gamma$) &      10.0 &          8.5 \\
\midrule
Er & 68 & 164 &               $<$0.5\% &                       0.09 &                       0.05 & $^{164}$Ho(n,$\gamma$) &     6.9 &          2.7 \\
Er & 68 & 166 &              23-63\% &                       1.00 &                       0.75 & $^{166}$Dy(n,$\gamma$) &       7.1 &          3.5 \\
Er & 68 & 167 &               3-23\% &                       0.86 &                       0.70 & $^{167}$Ho(n,$\gamma$) &       9.6 &          7.2 \\
Er & 68 & 168 &               9-32\% &                       0.45 &                       0.32 & $^{168}$Ho(n,$\gamma$) &       9.8 &          3.8 \\
\midrule
Hf & 72 & 176 &               $<$0.5\% &                       1.00 &                       0.86 & $^{175}$Yb(n,$\gamma$) &     8.7 &          5.6 \\
Hf & 72 & 177 &               5-29\% &                       0.99 &                       0.84 & $^{177}$Yb(n,$\gamma$) &       9.3 &          5.9 \\
Hf & 72 & 178 &              20-44\% &                       0.58 &                       0.48 & $^{178}$Yb(n,$\gamma$) &       8.1 &          3.8 \\
Hf & 72 & 179 &               6-22\% &                       0.64 &                       0.57 & $^{179}$Lu(n,$\gamma$) &       9.6 &          6.6 \\
Hf & 72 & 180 &              28-57\% &                       0.44 &                       0.36 & $^{180}$Lu(n,$\gamma$) &      10.3 &          3.7 \\
\midrule
Os & 76 & 188 &              30-68\% &                       1.17 &                       0.62 &  $^{188}$W(n,$\gamma$) &       8.6 &          3.2 \\
Os & 76 & 189 &               4-22\% &                       0.65 &                       0.59 & $^{189}$Re(n,$\gamma$) &       3.8 &          4.6 \\
Os & 76 & 190 &              10-33\% &                       0.54 &                       0.39 &  $^{190}$W(n,$\gamma$) &      11.1 &          3.7 \\
Os & 76 & 192 &              13-29\% &                       0.31 &                       0.29 &  $^{190}$W(n,$\gamma$) &      11.1 &          3.7 \\
\midrule
Pb & 82 & 204 &               $<$0.5\% &                       1.31 &                       1.55 & $^{203}$Hg(n,$\gamma$) &     6.3 &          9.8 \\
Pb & 82 & 206 &               4-28\% &                       0.66 &                       0.68 & $^{210}$Pb(n,$\gamma$) &       5.9 &         15.1 \\
Pb & 82 & 207 &               3-17\% &                       0.52 &                       0.51 & $^{211}$Bi(n,$\gamma$) &       8.1 &         13.5 \\
\bottomrule
\end{longtable}